\documentstyle[12pt]{article}
 \input epsf
\renewcommand{\theequation}{\arabic{section}.\arabic{equation}}

\newcommand{\bq}[1]{\begin{equation}{#1}\end{equation}}
\newcommand{\la}[1]{\label{#1}}
\newcommand{\eq}[1]{eq.~(\ref{#1})}
\newcommand{\eqs}[2]{eqs.~(\ref{#1}, \ref{#2})}

\newcommand{\ur}[1]{~(\ref{#1})}
\newcommand{\urs}[2]{~(\ref{#1},~\ref{#2})}
\newcommand{\urss}[3]{~(\ref{#1},~\ref{#2},~\ref{#3})}
\newcommand{\Eq}[1]{Eq.~(\ref{#1})}
\newcommand{\Eqs}[2]{Eqs.(\ref{#1},\ref{#2})}

\newcommand{\ra}[1]{(\ref{#1})}
  \def\beq{\begin{equation}}
  \def\eeq{\end{equation}}
  \def\beqr{\begin{eqnarray}}
  \def\eeqr{\end{eqnarray}}
  \def\nn{\nonumber\\}
 
 \def\Tr{\mbox{Tr}}
 \def\Sp{\mbox{Sp}}

 \def\bra#1{{\langle#1\vert}}
 \def\ket#1{{\vert#1\rangle}}
 
 \def\Dirac#1{#1\hskip-6pt/}
 \def\dd{\Dirac\partial}

\begin{document}
\rightline{RUB-TPII-9/96}
\vskip 2true cm
\begin{center}
\Large\bf Nucleon Parton Distributions at Low Normalization Point in the
Large $N_c$ Limit \end{center}
\vskip 1true cm
\begin{center} {\large D.~Diakonov, V.~Petrov, P.~Pobylitsa, M.~Polyakov \\
{\small\it Petersburg Nuclear Physics Institute, Gatchina,
St.Petersburg 188350, Russia} \\ and \\ C.~Weiss \\
{\small\it Institut f\"ur Theoretische Physik II,
Ruhr-Universit\"at Bochum, D-44780 Bochum, Germany}}
\end{center}

\vskip 1.5true cm

\abstract{
At large $N_c$ the nucleon can be viewed as a soliton of the effective
chiral lagrangian. This picture of nucleons allows
a consistent nonperturbative calculation of the leading-twist parton
distributions at a low normalization point. We derive general formulae for
the polarized and unpolarized distributions (singlet and non-singlet) in
the chiral quark-soliton model. The consistency of our approach is
demonstrated by checking the baryon number, isospin and total momentum sum
rules, as well as the Bjorken sum rule.  We present numerical estimates of
the quark and antiquark distributions and find reasonable agreement with
parametrizations of the data at a low normalization point. In particular,
we obtain a sizeable fraction of antiquarks, in agreement with the
phenomenological analysis.
 \newpage
\tableofcontents
\newpage

\section{Introduction}
\setcounter{equation}{0}

The distributions of quarks, antiquarks and gluons in nucleons, as
measured in the inclusive deep inelastic scattering of leptons,
provides us probably with the largest portion of quantitative
information about strong interactions.  Until now only the
{\em evolution} of the structure functions from a high value of $q^2$ to
even higher values, has been successfully compared with the data.
It is the field of the perturbative QCD, and its success has been,
historically, essential in establishing the validity of the QCD itself.
Unfortunately, the initial conditions for that evolution, namely the
leading-twist distributions at a relatively low normalization point, belong
to the field of the nonperturbative QCD, and the success here is still
rather modest.

In this paper we attempt to calculate parton distributions at a low
normalization point in the limit of large number of colours, $N_c
\rightarrow\infty$. Even though in reality $N_c=3$, the academic limit
of large $N_c$ is known to be a useful guideline. At large $N_c$ the
nucleon is heavy and can be viewed as a classical soliton of the
pion field \cite{Witten}. An example of the dynamical realization of
this idea is  given by the Skyrme model \cite{ANW}. However, the
Skyrme model is based on an unrealistic effective chiral lagrangian.
A far more realistic effective chiral lagrangian is given by the
functional integral over quarks in the background pion field
\cite{DE,DP}:

\[
\exp\left(iS_{\rm eff}[\pi(x)]\right)=
\int D\psi D\bar\psi \exp\left(i\int d^4x
\bar\psi(i\dd - MU^{\gamma_5})\psi\right),
\]
\beq
U=\exp\left(i\pi^a(x)\tau^a\right),\;\;\;\;\;
U^{\gamma_5}=\exp\left(i\pi^a(x)\tau^a\gamma_5\right)=
\frac{1+\gamma_5}2 U
+ \frac{1-\gamma_5}2 U^\dagger.
\la{FI}\eeq
Here $\psi$ is the quark field, $M$ is the effective quark mass
which is due to the spontaneous breakdown of chiral symmetry (generally
speaking, it is momentum-dependent) and $U$ is the $SU(2)$ chiral pion
field. The effective chiral action given by \eq{FI} is known to
contain automatically the Wess--Zumino term and the four-derivative
Gasser--Leutwyler terms, with correct coefficients. Therefore, at least the
first four terms of the gradient expansion of the effective chiral
lagrangian are correctly reproduced by \eq{FI}, and chiral symmetry
arguments do not leave much freedom to further modifications.
\Eq{FI} has been derived from the instanton model of the QCD vacuum
\cite{DP,DP1}, which provides a natural mechanism of chiral
symmetry breaking and enables one to express the dynamical mass $M$ and the
ultraviolet cutoff $\Lambda$ intrinsic in \eq{FI} through the
$\Lambda_{QCD}$ parameter. The effective chiral theory \ur{FI}
is valid for the values of the quark momenta up to the ultraviolet
cutoff $\Lambda$. Therefore, in using \eq{FI} we imply that we are
computing the parton distributions at the normalization point about
$\Lambda\approx 600$~MeV. It should be mentioned that \eq{FI} is
of a general nature: one need not believe in instantons
and still use \eq{FI}.

An immediate implication of this effective chiral theory is the
quark-soliton model for baryons of ref.~\cite{DPP}, which is in the
spirit of the earlier works~\cite{KaRiSo,BiBa} but without the vacuum
instability paradox noticed there. According to the model nucleons
can be viewed as $N_c$ (=3) ``valence" quarks bound by a self-consistent
hedgehog-like pion field (the ``soliton") whose energy coincides in fact
with the aggregate energy of quarks from the negative-energy Dirac
continuum. Similarly to the Skyrme model large $N_c$ are needed as an
algebraic parameter to justify the use of the mean-field approximation
(like one needs large $Z$ to justify the Thomas--Fermi atom), however the
$1/N_c$ corrections can be and in some cases are computed \cite{WY,Chr}.
The quark-soliton model of nucleons developed in ref.~\cite{DPP} includes a
collective-quantization procedure to deal with the rotational excitations
of the quark-pion soliton. (The quantization of the otherwise static
solution is necessary to obtain physical baryon states with definite
quantum numbers). It enables one to calculate the $N$ and $\Delta$
properties, such as formfactors, $\Delta -N$ splitting, magnetic moments,
axial constants, etc. For a review of baryon properties obtained from the
model see \cite{Go} and references therein. Until now the model lacks
explicit confinement (though probably it can be implemented along the lines
discussed in ref. \cite{D}), but it seems to be not so important for the
{\em ground} state nucleon.

Turning to the calculation of the nucleon structure functions we note
that the model possesses all features needed for a successful description
of the nucleon parton structure: it is an essentially quantum
field-theoretical relativistic model with explicit quark degrees of
freedom, which allows an unambiguous identification of quark as well as
antiquark distributions in the nucleon. This should be contrasted to the
Skyrme model where it is not too clear how to define quark and antiquark
distributions.  The advantage of the model can be also seen if one compares
it to any variant of the bag model. Unfortunately, the bag surface is
not described consistently in terms of fields. Ignoring the
"structure function of the surface" the bag models run into a violation
of general theorems requiring the complete account for all constituents
(see section 8). It should be added that all modern fits to the data
tend to include antiquarks and gluons at a low normalization
point, below $1\;GeV^2$ \cite{MRS,L,GRV}.

In this paper we develop the framework for calculating polarized and
unpolarized quark and antiquark distributions of the nucleon as described
by the effective chiral theory \ur{FI}. We check the validity of general
theorems, like the sum rules for the baryon number, isospin and the total
momentum carried by partons, as well as the Bjorken sum rule for the
polarized distributions. We also derive the expression for the Gottfried
sum rule and show that its r.h.s. is generally non-zero.

We show that, from the viewpoint of large $N_c$, all distributions can be
divided into ``large" and ``small" ones. We estimate numerically the ``large"
distributions: the singlet unpolarized distribution and the isovector
polarized one, for quarks and antiquarks separately~\footnote{A few
preliminary numerical results have been announced in ref.~\cite{D}.}.
The obtained distributions should be, in principle, used as initial
conditions for the standard perturbative evolution to higher values of
$q^2$ where one can compare them with the available data. Actually, in this
paper we compare our results with the parametrization of the data at a
low normalization point performed recently by Gl\"uck, Reya
{\em et al.}~\cite{GRV,GR2}.

\newpage

\section{From QCD to the effective chiral theory}
\setcounter{equation}{0}

\subsection{Light-cone representation for distribution functions}

The unpolarized quark distribution function of flavour $f$ inside a
nucleon with 4-moment\-um $P$, averaged over its spin, is given by the
following QCD equation (see e.g.~\cite{Collins-Soper-82}):

\beqr
&&
q_f(x,\mu) = \frac{1}{4\pi} \int\limits_{-\infty}^\infty dz^-
e^{ixp^+z^-}
\nn
&&
\cdot\bra{P}\bar\psi_f(0) \gamma^+ \mbox{P}\exp
\left\{-i g \int_0^z dz'{}^\alpha A_\alpha(z') \right\}
\psi_f(z) \ket{P}
\Bigr|_{z^+=0,\>z_\perp=0, \>\mu} \,,
\la{quark1}\eeqr
where $\psi_f$ are quark fields and $A_\alpha$ is the gluon field.
The antiquark distribution is

\beqr
&&
\bar q_f(x,\mu)= - \frac{1}{4\pi}
\int\limits_{-\infty}^\infty dz^-  e^{-ixp^+z^-}
\nn
&&
\cdot\bra{P}\bar\psi_f(0) \gamma^+
\mbox{P}\exp \left\{-i g \int_0^z dz'{}^\alpha A_\alpha(z') \right\}
\psi_f(z) \ket{P} \Bigr|_{z^+=0,\>z_\perp=0, \>\mu}.
\la{quark2}\eeqr
Here we use the light-cone coordinates
\bq{
z^{\pm} = \frac{z^0\pm z^3}{\sqrt 2}, \qquad
\gamma^{\pm} = \frac{\gamma^0 \pm \gamma^3}{\sqrt 2},
\la{cone-coordinates}
}
and the nucleon state is normalized by
\bq{
\langle P | P' \rangle = 2P^0 \delta^{3}({\bf P} - {\bf P}')
\la{relativistic-normalization}
}
where $P$ is the nucleon 4-momentum. Throughout the paper $x$ is the
Bjorken variable, $x=-q^2/(2P\cdot q)$, where $q$ is the 4-momentum
transfer to the nucleon.

The physical meaning of these equations is clear:
the parton model description of the nucleon is justified in the infinite
momentum frame; by using a backward Lorentz transformation
one can recover a nucleon at rest but the price is that one has to work
with the quark correlation functions on the light cone.
The transition from the infinite momentum frame
to the light-cone correlation function
over the nucleon at rest is helpful since in the large $N_c$ limit the
nucleon is a heavy particle, and it is convenient to work
in the nucleon rest frame.

The right hand sides of \eqs{quark1}{quark2} depend on the QCD
normalization scale parameter $\mu$. Actually, the non-local light-cone
products of fields in these equations should be considered as a formal
compact notation for an expansion in a series of local operators, each of
these operators being renormalized at the scale $\mu$. In contrast to QCD,
the effective chiral field theory is nonrenormalizable and
contains an explicit ultraviolet cutoff $\Lambda$ (not to be mixed up with
the $\Lambda_{QCD}$ parameter). In the instanton vacuum model this cutoff
appears as the inverse average instanton size:
$\Lambda\approx\bar\rho^{-1}\approx 600$~MeV \cite{DP1}. The results
obtained below refer thus to the low QCD scale $\mu$ of the order of
$600$~MeV.

We stress that we are computing the leading-twist distributions at a low
normalization point and not the structure functions, observable in principle
at low $q^2$. The former differ from the latter by higher-twist power
corrections which are large at low $q^2$. The distributions we are computing
can be directly used as initial conditions to the standard perturbative
evolution to higher $q^2$ where the power corrections are suppressed so
that the distributions become directly related to the observables.

\Eqs{quark1}{quark2} allow one to introduce a single distribution function
$q_f(x)$ defined in the interval $[-1,1]$ identifying it at negative $x$
with the antiquark distribution:

\bq{
q_f(x) =\left\{
\begin{array}{ll}
q_f(x),& x>0\,, \\
- \bar q_f(-x),& x<0 \,.
\end{array}
\right.
\la{q-q-qbar}
}
With this definition of $q_f(x)$
\eq{quark1} is valid in the interval $-1\leq x\leq 1$.

An alternative way to the structure functions is via their moments.
For example, one defines the moments of the {\em singlet} structure
function as

\beq
M_n = \int\limits_{-1}^1 dx\, x^{n-1} \sum\limits_f q_f(x)\,.
\la{mom1}\eeq
Introducing local quark and gluon twist-2 operators \cite{GW},

\beq
{\cal O}_n^{q,\mu_1\ldots\mu_n} =\frac12\, \frac{i^{n-1}}{n!}
\left[
\bar \psi \gamma^{\mu_1} D^{\mu_2}\ldots D^{\mu_n} \psi + \mbox{
permutations} -\mbox{ traces } \right],
\la{twist2q}\eeq
\beq
{\cal O}_n^{G,\mu_1\ldots\mu_n} = \frac{i^{n-2}}{n!}\,
\mbox{Tr}\,
\left[
G^{\mu_1\nu} D^{\mu_2}\ldots D^{\mu_{n-1}} G_\nu{}^{\mu_n}
+ \mbox{ permutations}
-\mbox{ traces }\right]\,,
\la{twist2g}\eeq
one can express the moments of the singlet structure function as
nucleon matrix elements of local operators:

\beq
M_n=\frac{i^{n-1}}{2}  M_N^{-n}
\langle P | \bar\psi_f
{v\hskip-6pt/} (v^\mu D_\mu)^{n-1}\psi_f |P \rangle,
\la{mom2}\eeq
where $v_\mu$ is a light-like vector,

\beq
v_\mu v^\mu=0, \qquad v_\mu P^\mu =M_N.
\la{llv}\eeq

When one starts to work with the effective low-energy theory \ur{FI}
all information about gluons is already lost. Therefore, in order to
recover the gluon distribution one has to go one step back, before the
derivation of \eq{FI} \cite{DP} and to rewrite the gluon operators in
terms of (possibly nonlocal) quark operators. Ref. \cite{DPW} explains
how it can be done, and certain gluon operators are expressed there
through quark fields only. Applying the method of ref. \cite{DPW}
to the twist-2 gluon operators \ur{twist2g} one observes, however, that
in the leading order in the instanton packing fraction,

\beq
f=\bar\rho^4\frac{N}{V} \ll 1,
\la{pf}\eeq
these operators are zero. The reason is the
$O(4)$ symmetry of the instanton field: after integration over
instanton orientations one can build the $\mu_1\ldots\mu_n$ tensor
only out of the Kronecker symbols, but it is impossible to get it
traceless. In order to obtain a non-zero result one has to go beyond
the zero-mode approximation of ref. \cite{DPW} and/or consider
contributions of many instantons. Both ways would lead to extra powers
of the packing fraction of instantons. Meanwhile, it is the smallness
of this packing fraction which is used in the derivation of \eq{FI}.
\Eq{FI} contains actually two dimensional parameters: the constituent
quark mass $M$ and the ultraviolet cutoff $\Lambda$; algebraically

\beq
\frac{M^2}{\Lambda^2} \sim f\ll 1.
\la{MkL}\eeq
Hence, to be consistent with the effective chiral lagrangian
\ur{FI}, one has to neglect the gluon operators, and to replace the
covariant derivatives in \eqs{twist2q}{mom2} by ordinary ones. That is
what we are going to do in the rest of the paper. It should be kept in
mind, however, that one {\em has} to introduce a finite cutoff
$\Lambda$ in order to make the nucleon mass and some of the structure
functions finite (fortunately, the potential divergences are but
logarithmical). In fact gluons are resident in the
"formfactor" of constituent quarks whose size is $\sim 1/\Lambda$.
At $\Lambda \rightarrow \infty$ the constituent quarks are point-like,
and there are no gluons. For the actual finite value of
$\Lambda \approx 600$~MeV gluons have to show up.
Unfortunately, the precise form of the gluon distribution depends,
in the language of the chiral model, on the details of the ultraviolet
regularization, therefore we shall not attempt to determine the gluon
distribution in this paper.

In the ``quarks-antiquarks only" approximation one is left
with the following expression for the quark (and antiquark)
distribution function $q_f(x)$ at $-1<x<1$:

\bq{
q_f(x)=  \frac{1}{4\pi}
\int\limits_{-\infty}^\infty dz^- e^{ixp^+z^-}
\bra{P} \bar\psi_f(0) \gamma^+ \psi_f(z) \ket{P}
\Bigr|_{z^+=0,\>z_\perp=0}\,.
\la{distribution-direct-definition-z}
}
According to \ra{cone-coordinates} one can write

\bq{
\bar\psi_f(0) \gamma^+ \psi_f(z)
= \psi_f^\dagger(0)\gamma^0 \gamma^+ \psi_f(z)
= \frac1{\sqrt 2}
\psi_f^\dagger(0)(1+\gamma^0 \gamma^3) \psi_f(z) \,.
}
It follows from the constraint $z^+=0$ in
\ra{distribution-direct-definition-z} that $z^3=-z^0$.
Taking the nucleon at rest we have $P^+=M_N/{\sqrt 2}$, where $M_N$ is
the nucleon mass. Therefore, the quark (and antiquark) distribution
function \ra{distribution-direct-definition-z} is

\bq{
q_f(x)
=  \frac{1}{4\pi}
\int\limits_{-\infty}^\infty dz^0  e^{ixM_Nz^0}
\bra{P} \psi_f^+(0) (1+\gamma^0\gamma^3) \psi_f(z) \ket{P}
\Bigr|_{z^3=-z^0,\>z_\perp=0} \,.
\la{main}
}

For the {\em polarized} distributions, $\Delta q_f(x)$, one has to
insert the $\gamma_5$ matrix in this equation (see,
e.g. \cite{Anselmino-PhRep}). In this paper we limit ourselves to two
flavours, $u$ and $d$, neglecting altogether the strange quarks, therefore
for both polarized and unpolarized distributions two cases have to be
distinguished: flavour singlet and flavour nonsinglet or isovector. For
isovector distributions one has to insert the $\tau^3$ matrix in \eq{main}.
All four distributions are considered in this paper.

Before turning to the calculation of these functions in the chiral
quark-soliton model we make a general remark about the quark distributions
in the limit $N_c\to \infty$.

In this limit the quark part of the nucleon momentum in the infinite
momentum frame is distributed among $O(N_c)$ quarks and antiquarks so that
each quark and antiquark carries $O(1/N_c)$ fraction of the nucleon
momentum. This means that the quark distribution function is concentrated
at $x \sim 1/N_c$. Keeping in mind that the total number of quarks minus
the number of antiquarks in the nucleon is $N_c$ we can write the baryon
number sum rule:

\begin{equation} \sum\limits_{f} \int\limits_{-1}^{1} dx q_f(x) =
N_c \,.
\la{B-number:sum-rule} \end{equation}
Comparing this sum rule with the fact that quark distributions
are concentrated at $x \sim 1/N_c$ we conclude that the singlet quark
distribution has the following form in the large $N_c$ limit:

\bq{
\sum\limits_{f}q_f(x) = N_c^2 \rho(N_c x)
\la{qf-rho-1}
}
where the function $\rho(y)$ is stable in the limit $N_c\to\infty$.
Similarly, the isovector polarized distribution,
$\Delta u(x)-\Delta d(x)$, is normalized, via the Bjorken sum rule, to
$g_A$ which, theoretically, is of the order of $N_c$. Therefore, this
distribution has also the form of \eq{qf-rho-1}. It should be contrasted to
the isovector unpolarized distribution, $u(x)-d(x)$, normalized to the
isospin which is $O(N_c^0)$. Therefore, it is $N_c$ times smaller than
in \eq{qf-rho-1}, and so is the singlet polarized distribution. In all four
cases the antiquarks follow the same pattern as they are given by the
same function as quarks but at $x<0$. Exactly this behaviour of the
distribution functions is obtained below.

\subsection{Chiral quark-soliton model of the nucleon}

Integrating out quarks in \ra{FI} one finds the effective chiral action,

\begin{equation}
S_{\rm eff}[\pi^a(x)]=-N_c\Sp\log D(U)\,,\;\;\;\;\;
D(U)= i\partial_0 - H(U),
\label{SeffU} \end{equation}
where $H(U)$ is the one-particle Dirac hamiltonian,

\beq
H(U) = - i\gamma^0\gamma^k \partial_k + M\gamma^0 U^{\gamma_5} \,,
\la{hU}\eeq
and $\mbox{Sp}\ldots$ is a functional trace.

For a given time-independent pion field $U=\exp(i\pi^a({\bf x})\tau^a)$
one can find the spectrum of the Dirac hamiltonian,

\begin{equation}
H\Phi_n = E_n \Phi_n.
\la{Dirac-equation}\end{equation}
It contains the upper and lower Dirac continua (distorted by the
presence of the external pion field), and, in principle, may contain
discrete bound-state level(s), if the pion field is strong enough. If
the pion field has the unity winding number, there is exactly one
bound-state level which travels all the way from the upper to the lower
Dirac continuum as one increases the spatial size of the pion field
from zero to infinity \cite{DPP}. One has to fill in this level to get
a non-zero baryon number state. Since the pion field is colour blind,
one can put $N_c$ quarks on that level in the antisymmetric state in
colour. We denote the energy of the discrete level as
$E_{\rm lev},\;\;-M\leq E_{\rm lev}\leq M$.

The limit of large $N_c$ allows to use the mean-field approximation
to find the nucleon mass -- similarly to the Thomas--Fermi model of
large $Z$ atoms. To get the nucleon mass one has to add up
$N_cE_{\rm lev}$
and the energy of the pion field. Since the effective chiral lagrangian
is given by the determinant \ur{SeffU} the energy of the pion field
coincides exactly with the aggregate energy of the lower Dirac
continuum, the free continuum subtracted. The self-consistent pion
field is thus found from the minimization of the functional \cite{DPP}

\begin{equation}
M_N =\min_U N_c\left\{E_{\rm lev}[U]
+ \sum_{E_n<0}(E_n[U]-E_n^{(0)})\right\}.
\la{nm}\end{equation}
From symmetry considerations one looks for the minimum in a hedgehog
ansatz:

\beq
U_c({\bf x})=\exp\left[i\pi^a({\bf x})\tau^a\right]
=\exp\left[i{(\bf n\cdot \tau}) P(r)\right],\;\;\;\;r=|{\bf x}|,\;\;\;\;
{\bf n}=\frac{{\bf x}}{r}
\la{hedge}\eeq
where $P(r)$ is called the profile of the soliton.

The minimum \ur{nm} is degenerate in respect to the translation of
the soliton in space and to the rotation of the soliton field in
ordinary and isospin spaces. For the hedgehog field \ur{hedge} the two
rotations are equivalent. Including slow rotations of the saddle-point
pion field and quantizing it gives rise to the quantum numbers of the
nucleon: its spin and isospin components \cite{ANW,DPP}.
In order to take into account the translational and rotational zero modes
one has to make a unitary rotation of the quark eigenfunctions and to shift
their centre,

\beq
\Phi_n({\bf x})\rightarrow R \Phi_n({\bf x-X}),
\la{transf}\eeq
and to make a projection to a concrete nucleon state under
consideration. The projection into a nucleon state with given momentum
${\bf P}$ is obtained by integrating over all shifts ${\bf X}$ of the
soliton,

\beq
\langle {\bf P^\prime}|\ldots|\ {\bf P}\rangle
=\int d^3{\bf X}\;e^{i({\bf P^\prime-P})\cdot{\bf X}}\ldots
\la{totmom}\eeq
The projection to a nucleon with given spin ($S_3$) and isospin
($T_3$) components is obtained by integrating over all spin-isospin
rotations $R$ of the soliton,

\beq
\langle S=T,S_3,T_3|\ldots| S=T,S_3,T_3\rangle
= \int dR\;\phi^{\dagger\;S=T}_{S_3T_3}(R)\ldots\phi^{S=T}_{S_3T_3}(R),
\la{spisosp}\eeq
where $\phi^{S=T}_{S_3T_3}(R)$ is the rotational wave function of the
nucleon given by the Wigner finite-rotation matrix \cite{DPP}:

\beq
\phi^{S=T}_{S_3T_3}(R)=\sqrt{2S+1}(-1)^{T+T_3}D^{S=T}_{-T_3,S_3}(R).
\la{Wigner}\eeq
The four nucleon states have $S=T=1/2$, with $S_3,T_3=\pm 1/2$.
[Taking the next rotational excitation with $S=T=3/2$ one can as
well compute the $\Delta$-resonance structure functions]. It is implied
that $dR$ in \eq{spisosp} is the Haar measure normalized to unity,
$\int dR =\int d(AR) = \int d(RB) = 1$. In what follows we shall
omit the superscript $S=T$.

\subsection{Light-cone quark correlation functions in the nucleon}

In this subsection we derive several equivalent ways of presenting the quark
distribution function \ur{main} in the chiral quark-soliton model.
Depending on the circumstances one can choose a more convenient
representation.

\Eq{main} is a matrix element of a non-local quark bilinear operator over
the nucleon state with definite 4-momentum $P$ and spin and isospin
components. According to the previous subsection one can write down a
general equation for such matrix elements; the time dependence of the
quark operators is accounted for by the energy exponents. We write
explicitly all the flavour ($f,g=1,2$) and the Dirac ($i,j=1,...,4$)
indices for clarity:

\beqr
&&
\bra{{\bf P},S_3,T_3}\psi^\dagger_{fi}(x^0,{\bf x})\psi^{gj}(y^0,{\bf y})
\ket{{\bf P},S_3,T_3}= 2P_0N_c \int d^3{\bf X}
\int dR\;\phi_{S_3T_3}^\dagger(R)
\nn
&&
\cdot
\sum\limits_{\scriptstyle n\atop \scriptstyle{\rm
occup.}}\exp[iE_n(x^0-y^0)]\Phi_{n,f^\prime i}^\dagger({\bf
x-X})R^{\dagger\,f^\prime}_fR_{g^\prime}^g\Phi_n^{g^\prime j}({\bf y-X})
\phi_{S_3T_3}(R)\,.
\la{annihilate}\eeqr
The functions $\Phi_n$ are eigenstates of energy $E_n$ of the Dirac
hamiltonian \ur{hU} in the external (self-consistent) pion field $U_c$.
Summation over colour indices is implied in the quark bilinears, hence the
factor $N_c$ in the r.h.s.

In \eq{annihilate} the quark is first annihilated in the nucleon by the
operator $\psi({\bf y})$ and only then created by the operator
$\psi^\dagger({\bf x})$. Therefore the sum goes over occupied states. For
the opposite ordering of the quark field operators, the order of creation
and annihilation is opposite and the sum runs over non-occupied states:

\beqr
&&
\bra{{\bf P},S_3,T_3}\psi^{gj}(y^0,{\bf y}) \psi^\dagger_{fi}(x^0,{\bf x})
\ket{{\bf P},S_3,T_3}= 2P_0N_c \int d^3{\bf X}
\int dR\;\phi_{S_3T_3}^\dagger(R)
\nn
&&
\cdot
\sum\limits_{\scriptstyle n\atop \scriptstyle
{\rm non-occup.}}\exp[iE_n(x^0-y^0)]\Phi_{n,f^\prime i}^\dagger({\bf
x-X})R^{\dagger\,f^\prime}_fR_{g^\prime}^g\Phi_n^{g^\prime j}({\bf y-X})
\phi_{S_3,T_3}(R)\,.
\la{create}\eeqr

We note that the eigenfunctions of the Dirac hamiltonian form a complete
set of functions only when both occupied and non-occupied states are
taken into account:

\beq
\sum\limits_{n\atop {\rm all}}\Phi_{n,fi}^\dagger({\bf x})
\Phi_n^{gj}({\bf y})
 =\delta_f^g\delta_i^j \delta({\bf x}- {\bf y}).
\la{commutator-static}\eeq

Adding up \eqs{annihilate}{create} at $x^0=y^0$ and using the completeness
condition \ur{commutator-static} we observe that these equations are
compatible with the standard equal-time anticommutator,

\beq
\left\{\psi_{fi}^\dagger({\bf x}),\psi^{gj}({\bf y})\right\}
=\delta_f^g\delta_i^j \delta({\bf x-y})\,.
\la{eqtac}\eeq

In a Lorentz-invariant field theory the fermion operators should
anticommute for any space-like separation:

\bq{ \left\{ \psi(y),
\psi^\dagger(x) \right\} = 0, \qquad\mbox{if } (x-y)^2 < 0 \,.
\la{locality-general} }

Our starting point, \eq{main}, is a nucleon matrix element of a product of
quark operators with a {\em light-like} separation which should be
understood as a limit of {\em space-like} separations. Indeed, the
light-like separation is obtained from the infinite momentum frame. As long
as the momentum is large but finite one has a space-like separation.
Therefore, the $\psi,\;\psi^\dagger$ operators in \eq{main} anticommute, and
two alternative though equivalent representations \urs{annihilate}{create}
can be written for the distribution functions; one sums over the occupied
states, the other sums over non-occupied states of the Dirac hamiltonian.

Finally, there exists another representation for the structure functions,
this time through the imaginary part of the Feynman Green function in the
background pion field. Let $G(\omega_1,{\bf p_1};\;\omega_2,{\bf p_2})$ be
the Fourier transform of the Feynman two-point Green function

\beqr &&
G(x^0, {\bf x};\; y^0, {\bf y})
= -\bra{y^0,{\bf y}}\frac{1}{i\Dirac{\partial}-MU^{\gamma_5}}
\ket{x^0,{\bf x}}
\nn
&&
= \bra{y^0,{\bf y}}(i\Dirac{\partial}+MU^{-\gamma_5})
\frac{1}{\partial^2+M^2-iM(\Dirac{\partial}U^{-\gamma_5})-i0}
\ket{x^0,{\bf x}},
\la{FGf}\eeqr
the free Green function being

\beq
G^{(0)}(p_1;p_2)
=(2\pi)^4\delta^{(4)}(p_1-p_2)\frac{M+\Dirac{p}_1}{M^2-p_1^2-i0}.
\la{freeG}\eeq

In principle, one can expand the Green function \ur{FGf} in powers
of the pion field, $U-1$, and its derivatives, $\partial U$. For the
time-independent pion field $U$ one has

\beq
G(p_1^0,{\bf p_1};\;p_2^0,{\bf p_2})=2\pi \delta(p_1^0-p_2^0)
S(p_1^0, {\bf p_1}, {\bf p_2}).
\la{t-ind}\eeq

Using this Green function we derive in Appendix A the following
representation for the singlet distribution function:

\beq
\sum\limits_{f} q_f(x) = -\mbox{ Im }\frac{N_cM_N}{2\pi}
\int\frac{d^4p}{(2\pi)^4}\;2\pi\delta(p^0+p^3-xM_N)\;
\Tr[S(p^0,{\bf p}, {\bf p})(\gamma^0+\gamma^3)].
\la{ImGF}\eeq

Since the Green function $S(p^0, {\bf p},{\bf p})$ can be expanded in the
pion field and/or its derivatives one can use this representation to
make a quick estimate of the structure function, see section~7.

\newpage

\section{Singlet unpolarized distribution}
\setcounter{equation}{0}

\subsection{Sum over quark levels}

We start with the flavour singlet distribution function,
$u(x)+d(x)$, and the antiquark one, $\bar u(x)+\bar d(x)$.
The singlet case is simpler than the isovector one since
one can ignore the rotation of the soliton. Indeed in that case
the indices of the orientation matrices in \eq{annihilate}
are contracted, $R^\dagger R=1$, and the integral over orientations $R$ in
\eq{annihilate} becomes trivial and equal to unity. Taking
the nucleon at rest, ${\bf P}=0,\;P_0=M_N$, and using \eq{annihilate} for
a specific light-cone separation of quarks operator as suggested by
\eq{main} we get:

\beqr
&&
u(x)+d(x) =  \frac{N_c
M_N}{2\pi}  \int d^3{\bf X} \int\limits_{-\infty}^\infty dz^0
e^{ixM_Nz^0} \sum\limits_{\scriptstyle n\atop \scriptstyle {\rm
occup.}} e^{-i E_n z^0}
\nn
&&
\cdot \Phi_{n,fi}^\dagger(-{\bf X})
(1+\gamma^0\gamma^3)^i_j
\Phi_{n}^{fj}({\bf z}-{\bf X})
\Bigr|_{z^3=-z^0,\>z_\perp=0} \,.
\la{singl1}\eeqr
We remind the reader that the same equation gives, at $x<0$, the
antiquark distribution, $\bar u(x) +\bar d(x)=-(u(-x)+d(-x))$.
Passing to the momentum representation,

\beq
\Phi_n({\bf x}) = \int \frac{d^3{\bf p}}{(2\pi)^3}\Phi_n({\bf p})
e^{i({\bf p\cdot x})},
\la{Fou}\eeq
and integrating over the coordinates of the nucleon center of inertia,
${\bf X}$, we obtain:

\beqr
&&
u(x)+d(x)=  \frac{N_c M_N}{2\pi}\!
\int\limits_{-\infty}^\infty\! dz^0  e^{ixM_Nz^0}\!\!
\sum\limits_{\scriptstyle n\atop \scriptstyle{\rm occup.}}\!
e^{-i E_n z^0}
\bra n(1+\gamma^0\gamma^3) \exp(-iz^0 p^3)\ket n
\nn
&&
=N_cM_N\!\!\sum\limits_{\scriptstyle n\atop \scriptstyle{\rm occup.}}\!
\bra n(1+\gamma^0\gamma^3)\delta(E_n+p^3-xM_N)\ket n \,,\;\;\;\;
x\in[-1,1],
\la{distribution-occupied}
\eeqr
where $\ket{n}=\Phi_n({\bf p})$; taking the matrix element implies
integration over $d^3{\bf p}/(2\pi)^3$. However, \eq{distribution-occupied}
can be understood in any other basis provided $p^k= - i\partial/\partial
x^k$ is the momentum operator.

An alternative representation for the distribution function arises
from \eq{create} where summation goes over {\em non-occupied} states:

\beqr
&&
u(x)+d(x)
= - \frac{N_c M_N}{2\pi}\!
\int\limits_{-\infty}^\infty\! dz^0  e^{ixM_Nz^0}
\!\!\!\!\sum\limits_{\scriptstyle n\atop \scriptstyle\!\!\!\!
{\rm non-occup.}}e^{-i E_n z^0}
\bra n (1+\gamma^0\gamma^3) \exp(-iz^0 p^3)\ket n
\nn
&&
=-N_cM_N\!\!
\sum\limits_{\scriptstyle n\atop \scriptstyle{\rm non-occup.}}\!\!
\bra n (1+\gamma^0\gamma^3)\delta(E_n+p^3-xM_N)\ket n \,,\;\;\;\;
x\in[-1,1].
\la{distribution-non-occupied}\eeqr

Both representations, over occupied and non-occupied states of the
Dirac hamiltonian, should be absolutely equivalent were the theory
finite. However, the effective chiral action \ur{SeffU} implies
an ultraviolet cutoff: the nucleon mass itself and its structure
functions diverge logarithmically. In order to support the equivalence
of two representations, \eq{distribution-occupied} and
\eq{distribution-non-occupied}, the ultraviolet regularization should
be introduced in such a way as to preserve the anticommutativity
of the quark fields at a space-like separation. In particular,
it means that the completeness of the fermion states, leading to the
equal-time anticommutator \ur{commutator-static} must not be violated.
For example, a relativistically-invariant regularization by the
Pauli--Villars method preserves the equivalence of
\eqs{distribution-occupied}{distribution-non-occupied}
and other necessary properties of the distribution functions.
On the contrary, a naive energy cutoff in the Dirac
continua violates the completeness, and is hence unacceptable.

In all above formulae we implied the vacuum
subtraction: whenever we write
$\bra {N} A \ket {N}$
with a local operator $A$, we mean in fact
$\bra {N} A - \bra 0 A \ket 0 \ket {N}$.
Without this subtraction $\bra {N} A \ket {N}$ contains
a divergence proportional to $\delta^{(3)}(0)$. The vacuum subtraction is
effectively equivalent to the subtraction in
\eqs{distribution-occupied}{distribution-non-occupied} of the appropriate
sums over eigenstates $\ket{n^{(0)}}$ with eigenenergies $E_n^{(0)}$ of the
free Dirac operator $H_0$.

For the free Dirac hamiltonian one has obviously
$E_n^{(0)} = \pm \sqrt{|{\bf p}_n|^2+M^2}$.
Therefore the delta function $\delta(xM_N- E_n^{(0)} - p_n^3)$ vanishes
if $\mbox{sign }(x)=-\mbox{sign} E_n^{(0)}$. Hence in
\eq{distribution-occupied} the vacuum subtraction term vanishes
at $x>0$ (that is for quarks) whereas in \eq{distribution-non-occupied}
the vacuum subtraction is not needed at $x<0$, that is for antiquarks.

Both quark and antiquark distribution functions must be positive.
At $x>0$ the function $u(x)+d(x)$ is given by
\eq{distribution-occupied} without vacuum subtraction.
We can rewrite this equation as

\beq
u(x)+d(x)= \frac{N_c M_N}{2\pi} \int \frac{d^2p_\perp}{(2\pi)^2}
\sum\limits_{\scriptstyle n\atop \scriptstyle
{\rm occup.}}\left[\Phi_n^\dagger({\bf p}) (1+\gamma^0\gamma^3)
\Phi_n({\bf p})\right]_{p^3=xM_N- E_n},\;\;\;\;x>0.
\la{distribution-occupied-positive}
\eeq
This expression is explicitly positive because $(1+\gamma^0\gamma^3)/2$
is an orthogonal projector.

Similarly, for the antiquark distribution one can write an
explicitly positive expression:
\beqr
&&
\bar u(x)+\bar d(x)=-[u(-x)+d(-x)]
\nn
&&
= \frac{N_c M_N}{2\pi} \int \frac{d^2p_\perp}{(2\pi)^2}\!\!\!
\sum\limits_{\scriptstyle n\atop \scriptstyle
{\rm non-occup.}}\!\!\!
\left[\Phi_n^\dagger({\bf p})(1+\gamma^0\gamma^3)\Phi_n({\bf p})
\right]_{p^3=-xM_N- E_n},\;\;\;\; x>0 \,.
\la{antiquark-distribution-occupied-positive}
\eeqr

Although these formulae are explicitly positive one should keep in mind
that they are ultraviolet divergent. In principle, one can regularize them
so that the positivity is preserved. However it is not obvious
{\em a priori} that one can find a regularization that simultaneously
preserves both positivity and the equivalence of
\eq{distribution-occupied} with \eq{distribution-non-occupied}.
We have observed that the Pauli--Villars method is favoured in this
respect, too.

\subsection{Trace representation for distribution functions}

All the above representations for quark distribution functions are written
as sums of diagonal matrix elements of certain operators over either
occupied or non-occupied eigenstates of the one-particle Dirac hamiltonian
$H$. Therefore one can easily rewrite these sums as operator traces.  To
this end we apply to \eq{distribution-occupied}
the spectral decomposition

\bq{
\sum\limits_{\scriptstyle n\atop \scriptstyle
{\rm occup.}} \ket n \bra n e^{-iz^0 E_n}
= \int\limits_{-\infty}^{ E_{\rm lev}+0} d\omega\;\delta(\omega-H)
e^{-iz^0 \omega}
}
and obtain

\beq
u(x)+d(x) =  N_c M_N\int\limits_{-\infty}^{ E_{\rm lev}+0} d\omega\;
\Sp\left[ \delta(\omega-H) \delta(\omega+p^3-xM_N)
(1+\gamma^0\gamma^3) \right]-(H\rightarrow H_0)\,,
\la{distribution-occupied-delta}
\eeq
where $\Sp\ldots$ is the functional trace.

Similarly, starting from the quark distribution function
in the form of the sum over non-occupied states,
\eq{distribution-non-occupied},
we arrive to the following representation:
\beq
u(x)+d(x) = -  N_c M_N\int\limits_{ E_{\rm lev}+0}^{+\infty}\!\!d\omega\;
\Sp\left[\delta(\omega-H)\delta(\omega+p^3-xM_N)(1+\gamma^0\gamma^3)
\right]-(H\rightarrow H_0)\,.
\la{distribution-non-occupied-delta}\eeq
Both formulae are valid at $-1<x<1$.
We remind the reader that, according to the results of the previous
subsection, in \eq{distribution-occupied-delta} one has to make a vacuum
subtraction at $x<0$ whereas in \eq{distribution-non-occupied-delta}
the vacuum subtraction is needed at $x>0$.

Symbolical as they may seem, \eqs{distribution-occupied-delta}
{distribution-non-occupied-delta} give a practical way of computing the
structure functions. To saturate the functional trace one can use
any complete set of functions. For example, the quark eigenfunctions of
the Dirac hamiltonian \ur{Dirac-equation} may be used. In this
basis the hamiltonian $H$ is diagonal but the momentum $p^3$ is
not. Another way is to use the eigenfunctions of the free hamiltonian,
the so-called Kahana--Ripka basis \cite{KR}. The trace representation is
also helpful in deriving general relations in a laconic form, see below.

\subsection{Baryon number sum rule}

Let us show that the baryon number sum rule is automatically satisfied
in the above equations. To get the baryon number sum rule one has to
integrate \ra{distribution-occupied-delta} over $x$ from $-1$ to $1$.
In the r.h.s. of that equation $x$ enters through the product $xM_N$
where $M_N=O(N_c)$, so that in the large $N_c$ limit we can replace
this integral by the integral over the whole real axis of $x$,
which leads to the following result:

\beq \int\limits_{-1}^{1} dx \, [u(x)+d(x)] =
N_c\int\limits_{-\infty}^{ E_{\rm lev}+0} d\omega\;\Sp\left[
\delta(\omega-H) (1+\gamma^0\gamma^3) \right] - (H\to H_0) \,.
\la{interm}\eeq
Owing to the rotational hedgehog symmetry of the soliton the term
$\gamma^0\gamma^3$ gives no contribution, and we are left with

\bq
{\int\limits_{-1}^{1} dx\,[u(x)+d(x)] =  N_c
\;\Sp\left[ \theta(-H +E_{\rm lev}+0) - \theta(-H_0)
\right],
}
which counts the number of the filled levels of the Dirac
hamiltonian, the number of the levels in the free lower Dirac
continuum subtracted. According to ref. \cite{DPP} it is the baryon
number of the state. We have thus proved the sum rule

\bq{
\int\limits_{-1}^{1} dx \, [u(x)+d(x)]
=\int\limits_0^{1} dx [u(x)+d(x) - \bar u(x)-\bar d(x)]
=  N_cB
\la{baryon-number:sum-rule}
}
where $B$ is the baryon number of the state; $B=1$ for the nucleon.

\subsection{Moments of distribution functions}

We define the moments of the singlet structure function as

\bq{M_n = \int\limits_{-1}^1 dx\, x^{n-1} \sum\limits_f q_f(x)\,.
\la{M-n-definition}
}

Let us multiply \ra{distribution-occupied-delta} by $x^{n-1}$
and integrate over $x$. In the large $N_c$ limit one can extend the
integration region to $-\infty < x < \infty$. Integrating the quark
distribution in the form of \eq{distribution-occupied-delta}
we obtain the following representation for the moments:

\beq
M_n =  N_c M_N^{1-n}\int\limits_{-\infty}^{ E_{\rm lev}+0} d\omega\;
\Sp\left[ \delta(\omega-H) (\omega+p^3)^{n-1}
(1+\gamma^0\gamma^3) \right]
- (H\to H_0)\,.
\la{Moment-occupied}\eeq
Note that we need the vacuum subtraction here since this result is
derived by integrating \ra{distribution-occupied-delta} over both positive
and negative $x$'s.

Similarly, the representation \ra{distribution-non-occupied-delta}
based on the summation over non-occupied states leads to
the alternative expression for the moments:

\beq
M_n = - N_c M_N^{1-n}\int\limits_{ E_{\rm lev}+0}^{+\infty} d\omega\;
\Sp\left[ \delta(\omega-H) (\omega+p^3)^{n-1}
(1+\gamma^0\gamma^3) \right]
- (H\to H_0).
\la{Moment-non-occupied}\eeq

Both representations in fact follow from a third representation
where the integration over $\omega$ goes along the imaginary axis, to the
right of $E_{\rm lev}$. Putting $\omega = i\omega^\prime$ we can write

\beq
M_n = -i N_c M_N^{1-n}\int\limits_{-\infty}^{+\infty}
\frac{d\omega^\prime}{2\pi}
\Sp\left[\frac{1}{\omega^\prime+iH} (i\omega^\prime+p^3)^{n-1}
(1+\gamma^0\gamma^3) \right]
- (H\to H_0)\,.
\la{Moment-contour}\eeq
Indeed, closing the $\omega^\prime$ integration contour
to the upper cut (corresponding to the lower Dirac continuum
plus the discrete level) or to the lower cut and pole
(corresponding to the upper Dirac continuum) one immediately reproduces
\eqs{Moment-occupied}{Moment-non-occupied}. Note that we assume that
the $\omega^\prime$ integration contour in \eq{Moment-contour} is
chosen so that the discrete level belongs to the occupied states.

The fact that different deformations of the $\omega^\prime$
integration contour in \eq{Moment-contour} lead to the two
representations in terms of summation over occupied and non-occupied
states respectively, gives another proof of the equivalence of these
equations. We remind the reader that earlier we have derived this
equivalence from causality, see subsection 2.3. It illustrates a
general connection between causality and analyticity.

However, an arbitrary ultraviolet regularization of the theory may
easily destroy both causality and analyticity. One has to check that
a particular regularization does not induce new singularities in the
$\omega^\prime$ plane, and that it does not prevent one from closing
the contours to the upper or to the lower cuts. For example, the popular
proper time regularization would violate the last requirement. On the
contrary, the Pauli-Villars regularization does not spoil the analytic
properties, as well as casuality, see subsection 3.1.

\subsection{Momentum sum rule}
\la{Momentum-sum-rule-section}

Let us derive the momentum sum rule for the quark distribution
functions. It holds true only if equations of motion are satisfied,
i.e. the pion field $U$ which binds quarks in the nucleon is the
minimum of the functional \ur{nm} which can be symbolically written
as

\beq
M_N=N_c\;\Sp\left[\theta( E_{\rm lev} -H + 0)H \right]- (H\to H_0).
\la{nm1}\eeq
The saddle-point equation for the self-consistent pion field reads:

\beq
\Sp \bigl[ \theta( E_{\rm lev} +0 -H) \delta_U H \bigr] =0\,.
\la{saddle-point-equation}\eeq
where $\delta_U H$ is an arbitrary variation of the Dirac hamiltonian
under a variation of the chiral field $U$. Let us consider a
particular (dilatational) variation of the chiral field, $U(x)  \to
U[(1+\xi)x]$ with an infinitesimal $\xi$ such that $\delta U = \xi x^k
\partial_k U$. Then the correspondent variation of the Dirac
Hamiltonian is

\[
\delta_U H = M\gamma^0 \xi x^k \partial_k U^{\gamma_5}
= \xi   [ x^k  \partial_k,  M\gamma^0
U^{\gamma_5}] = \xi \left(  [ x^k  \partial_k,  H]
-i\gamma^0\gamma^k\partial_k \right).
\]
Inserting this variation into the saddle-point equation
\ra{saddle-point-equation} and taking into account that

\[
\Sp \bigl( \theta( E_{\rm lev}+0 -H) [ x^k
\partial_k,  H] \bigr) =\Sp \bigl( [H,\theta( E_{\rm lev} +0 -H) ] x^k
\partial_k \bigr) = 0,
\]
we get a useful identity:

\bq{
\Sp \bigl( \theta( E_{\rm lev} +0 -H)
\gamma^0\gamma_k\partial_k \bigr) = 0\,.
\la{useful-identity-1}
}
Owing to the hedgehog symmetry of the self-consistent pion field
one can write a general tensor

\bq{
\Sp \bigl( \theta( E_{\rm lev} +0 -H)
\gamma^0\gamma_k\partial_l \bigr) = \delta_{kl}A
\la{useful-identity-2}
}
with the zero constant $A$ because of \eq{useful-identity-1}.

Having made the necessary preparations, we turn to the second moment of
the structure function $M_2$.  Using the representation
\ur{Moment-occupied} for the moments we can write

\beqr &&
M_2=N_c M_N^{-1} \int\limits_{-\infty}^{ E_{\rm lev}+0} d\omega\;\Sp\left[
\delta(\omega-H) (\omega+p^3) (1+\gamma^0\gamma^3) \right] - (H\to H_0)
\nn &&
=  N_c M_N^{-1} \Sp  \left[ \theta( E_{\rm lev} -H + 0) (H+p^3)
(1+\gamma^0\gamma^3) \right] - (H\to H_0)
\nn &&
=  N_c M_N^{-1} \Sp\left[ \theta( E_{\rm lev} -H + 0)
(H+ p^3\gamma^0\gamma^3) \right] - (H\to H_0) \,.
\la{M-2}\eeqr
At the last step we have omitted the terms which do not contribute
to the trace owing to the hedgehog symmetry of the saddle point chiral
field. The identity \ra{useful-identity-1} allows us to ignore the
$p^3\gamma^0\gamma^3$ piece. Comparing \eq{M-2} with the
expression for the nucleon mass, \eq{nm1}, we obtain finally:

\bq{
M_2\equiv\int\limits_0^1dx\>x [u(x)+d(x)+\bar u(x)+\bar d(x)] =1,
\la{momentum-sum-rule-nonregularized}
}
meaning that quarks and antiquarks do carry the total momentum of the
nucleon. The Pauli--Villars regularization is again privileged in that
it does not destroy the momentum sum rule.

\newpage

\section{Isovector unpolarized distribution}
\setcounter{equation}{0}

\subsection{Soliton rotation}

It is easy to see that the isovector quark distribution function,
$u(x)-d(x)$, vanishes in the leading order of the $1/N_c$ expansion.
In order to obtain a nonvanishing result one has to consider rotational
corrections to the classical soliton. In the leading order of the
$1/N_c$ expansion the rotation of the soliton is taken into account by
the rotational wave functions $\phi_{S_3T_3}(R)$ -- see
\eq{Wigner}. In higher orders in $1/N_c$ one has to take into account
that the functional integral \ur{FI} goes over a time-dependent chiral
field,

\beq
U(t,{\bf x}) = R(t) U_c({\bf x}) R^\dagger(t)
\la{rot}\eeq
For the rotating ansatz we have

\beq
i\partial_t - H(U) =
R\;[i\partial_t - H(U_c) + i R^\dagger \dot R ]\;R^\dagger.
\la{rotH}\eeq
This leads to the following modification of the leading-order
\eqs{annihilate}{t-ind}, which takes into account the time dependence of the
Green function \ur{FGf}:

\beqr
&&
\bra {{\bf P}=0,S=T,S_3,T_3}
\psi^\dagger(x) \Gamma \psi(y)  \ket {{\bf P}=0,S=T,S_3,T_3}
\nn
&&
=2M_N i \int d^3{\bf X} \int dR \, \phi_{S_3T_3}^\dagger(R)
\nn
&&
\cdot \Tr \left\{ R^\dagger \Gamma R \bra{y^0,{\bf y}-{\bf X}}
[i\partial_t - H(U_c) + i R^\dagger \dot R ]^{-1}
\ket{x^0,{\bf x}-{\bf X}}  \right\}\phi_{S_3T_3}(R),
\la{time-annihilate-X-R-correction}
\eeqr
where $\Gamma$ is an arbitrary matrix in spin and isospin.
$\Tr\ldots$ denotes an ordinary trace in isospin and Lorentz
indices, as contrasted to $\Sp\ldots$ standing for functional traces;

The integrand should be now expanded in the angular
velocity $R^\dagger \dot R$ which should be replaced by the spin
operator $S$ according to the following quantization rule \cite{ANW,DPP}:

\beq
R^\dagger \dot R \to \frac{i}{2I} S^a\tau^a,
\la{quantization-rule}\eeq
where $I= O(N_c)$ is the moment of inertia of the soliton, see below,
\eq{momin}. At large $N_c$ the moment of inertia is large, and the
rotation may be considered as slow.

\subsection{Expressing the isovector distribution through a double sum
over levels}

In the leading order in $1/N_c$ one can simply neglect $R^\dagger \dot
R$ in \eq{time-annihilate-X-R-correction}. In this approximation
one reproduces \eq{singl1}. However, neglecting the angular velocity
altogether makes the nonsinglet distribution vanish. The first
nonvanishing contribution to the isovector unpolarized quark
distribution comes from the term linear in $R^\dagger \dot R$ in
\ra{time-annihilate-X-R-correction}.  Inserting this term into the
general formula for the quark distribution function
\ur{main} we obtain

\begin{eqnarray} &&
u(x) - d(x) =
\frac{iM_N N_c}{4\pi} \sum\limits_{S_3}
\int\limits_{-\infty}^\infty dz^0  e^{ixM_Nz^0}
\int d^3{\bf X} \int dR \, \phi_{S_3T_3}^\dagger(R)
\nonumber\\
&&
\cdot
\Tr \Bigl\{ R^\dagger \tau^3 R(1+\gamma^0\gamma^3)\bra{z^0,{\bf z}-{\bf X}}
\frac1{i\partial_t - H(U_c)}(- i R^\dagger \dot R )
\nn
&&
\cdot
\frac1{i\partial_t - H(U_c)}\ket{0,-{\bf X}}  \Bigr\}
\Bigr|_{z^3=-z^0,\>z_\perp=0}\phi_{S_3T_3}(R) \,.
\la{triunpol1}\end{eqnarray}
Applying the quantization rule \ra{quantization-rule} and
introducing the orientation matrix in the adjoint representation,

\begin{equation}
D_{ab}(R) = \frac{1}{2} \Tr(R^\dagger\tau^a R \tau^b),
\la{D-definition}
\end{equation}
we find

\begin{eqnarray}
&&
u(x) - d(x) =
\frac{N_cM_N i}{8\pi I} \sum\limits_{S_3}
\int\limits_{-\infty}^\infty dz^0  e^{ixM_Nz^0}
\int dR \, \phi_{S_3T_3}^\dagger(R)
D_{3b}(R) S^a \phi_{S_3T_3}(R) \int d^3{\bf X}
\nonumber\\
&&
\cdot
\Tr \left\{( \tau^b (1+\gamma^0\gamma^3)
\bra{z^0,{\bf z}-{\bf X}}
\frac1{i\partial_t - H(U_c)}\, \tau^a
\frac1{i\partial_t - H(U_c)}
\ket{0,-{\bf X}}  \right\}
\Bigr|_{z^3=-z^0,\>z_\perp=0}
\,.
\end{eqnarray}
Let us first compute the rotational matrix element. Strictly speaking, this
matrix element contains noncommuting operators $D_{3b}(R)$ and $S^a$,
and one may worry about their ordering. However, due to the
summing over nucleon spin $S_3$ the result does not depend on the order:

\begin{eqnarray}
&&
\sum\limits_{S_3}
\int dR \, \phi_{S_3T_3}^\dagger(R)
D_{3b}(R) S^a \phi_{S_3T_3}(R)
\nonumber\\
&&
= \sum\limits_{S_3} \int dR \, \phi_{S_3T_3}^\dagger(R)
S^a D_{3b}(R)  \phi_{S_3T_3}(R)
= - \frac{1}{3} \delta^{ab} (2T^3)
\la{rotme}\end{eqnarray}
Therefore we get:

\begin{eqnarray}
&&
u(x) - d(x) = - (2T_3) \frac{N_cM_N i}{24\pi I}
\int\limits_{-\infty}^\infty dz^0  e^{ixM_Nz^0}\int d^3{\bf X}
\sum\limits_{a=1}^3 \Tr \Bigl\{ \tau^a(1+\gamma^0\gamma^3)
\nonumber\\
&&
\cdot\bra{z^0,{\bf z}-{\bf X}}\frac1{i\partial_t - H(U_c)}\, \tau^a
\frac1{i\partial_t - H(U_c)}\ket{0,-{\bf X}}  \Bigr\}
\Bigr|_{z^3=-z^0,\>z_\perp=0}\,.
\la{nonsinglet-distribution-result}
\end{eqnarray}
Since $H(U_c)$ is time-independent (it coincides now with the hamiltonian
$H$ of the previous sections) we rewrite the matrix element as

\begin{eqnarray}
&&
\bra{z^0,{\bf z}-{\bf X}}\frac1{i\partial_t - H}\,\tau^a
\frac1{i\partial_t - H}\ket{0,-{\bf X}}
\nonumber\\
&&
= \int\limits_{-\infty}^{\infty} \frac{d\omega}{2\pi}
e^{-i\omega z^0}\bra{{\bf z}-{\bf X}}
\frac1{\omega - H}\,\tau^a \frac1{\omega - H}\ket{-{\bf X}}\,.
\end{eqnarray}
Noticing that
$\bra{{\bf z}-{\bf X}}=\bra{-{\bf X}} \exp[i({\bf p\cdot z})]$
and integrating first over ${\bf X}$ and then over $z^0$ we get finally

\begin{eqnarray}
&&
u(x) - d(x) =- (2T_3) \frac{N_cM_N i}{24\pi I}
\int\limits_{-\infty}^{\infty} d\omega
\nonumber\\
&&
\cdot
\sum\limits_{a=1}^3\Sp \Biggl[\tau^a(1+\gamma^0\gamma^3)
\delta(\omega+p^3 - xM_N)
\frac1{\omega - H}\, \tau^a
\frac1{\omega - H}\Biggr]\,.
\la{triunpf}\end{eqnarray}

The result is of the same form as for the singlet structure function
\ur{distribution-occupied-delta}. However, in contrast to the singlet
distribution which is a {\em single} sum over occupied (or non-occupied)
levels (see \eqs{distribution-occupied}{distribution-non-occupied}),
the isovector distribution is a {\em double} sum over levels.
To see that explicitly we saturate the functional trace in \eq{triunpf}
by a complete set of functions, say, by the eigenfunctions of the Dirac
hamiltonian $\Phi_n({\bf p})$ in the momentum representation. Then
the integration over $\omega$ is performed with the help of the
$\delta$-function, and we get:

\begin{eqnarray}
&&
u(x) - d(x) = - (2T_3) \frac{N_cM_N i}{24\pi I}
\sum\limits_{a=1}^3 \sum\limits_{m,n} \int \frac{d^3{\bf p}}
{(2\pi)^3} \frac{\Phi_n^\dagger({\bf p}) \tau^a(1+\gamma^0\gamma^3)
\Phi_m({\bf p})}{(p^3 - xM_N - E_m)(p^3 - xM_N - E_n)}
\nonumber\\
&&
\cdot
\int \frac{d^3{\bf p^\prime}}{(2\pi)^3}
\Phi_n^\dagger({\bf p^\prime})\tau^a\Phi_m({\bf p^\prime})\,.
\la{isofin}\end{eqnarray}
The denominators here should be in fact understood with an $i\epsilon$
prescription following from the original
\eq{time-annihilate-X-R-correction}:

\begin{eqnarray}
E_m \to E_m + i0 & \qquad\mbox{for occupied states},
\nonumber\\
E_m \to E_m - i0 & \qquad\mbox{for non-occupied states}.
\la{ieps}\end{eqnarray}

We note that the wave functions $\Phi_n({\bf p})$ generically have
singularities in the complex plane, therefore one cannot, generally
speaking, close the contour of integration in $p^3$ in \eq{isofin}
and reduce the integral to a residue of one of the denominators in
\eq{isofin}~\footnote{In that way one would recover eq.(25) from a recent
paper \cite{WGR}. We conclude that their eq.(25) does not agree with
our result, \eq{isofin}.}.

\subsection{Isospin sum rule}

Let us now check the isospin sum rule,

\begin{equation}
\int\limits_{0}^{1}dx\{u(x)-d(x)-[\bar u(x)-\bar
d(x)]\}\equiv\int\limits_{-1}^{1} dx [u(x) - d(x)] = 2T_3.
\la{nonsinglet-sum-rule} \end{equation}
As usually for general relations, the derivation of this sum rule is more
laconic when one uses the symbolic expressions with functional traces.
Indeed, integrating \eq{triunpf} over $x$ and replacing in the large $N_c$
limit the integration limits $[-1,1]$ by the whole real $x$ axis  we
obtain:

\beq
\int\limits_{-1}^{1} dx [u(x) - d(x)]
= - (2T_3) \frac{iN_c}{12 I}
\int \frac{d\omega}{2\pi}  \sum\limits_{a=1}^3
\Sp \Bigl( \tau^a (1+\gamma^0\gamma^3)
\frac1{\omega - H}\, \tau^a \frac1{\omega - H}\Bigr) \,.
\eeq
The hedgehog symmetry allows us to drop the term $\gamma^0\gamma^3$.
Taking into account that the nucleon moment of inertia
is~\cite{DPP}

\begin{equation}
I = -  \frac{iN_c}{12}
\int \frac{d\omega}{2\pi}  \sum\limits_{a=1}^3
\Sp \Bigl( \tau^a \frac1{\omega - H}\, \tau^a
\frac1{\omega - H} \Bigr),
\la{momin}\end{equation}
we immediately reproduce the sum rule \ra{nonsinglet-sum-rule}.
One should keep in mind that both the moment of inertia and
the isovector quark distribution contain logarithmic divergences,
so that their regularization should be consistent with one another, if
one wants to preserve the isospin sum rule.

\subsection{Gottfried sum rule}

The Gottfried sum rule \cite{Gottfried} follows from the assumption
that the antiquark distribution in the nucleon is isotopically
invariant:

\begin{equation}
\int_0^1 dx [\bar u(x) - \bar d(x)] =
-\int_{-1}^0 dx [u(x) - d(x)]=0\,.
\la{Go-sum-rule-definition}
\end{equation}

This relation does not follow from any fundamental principle of QCD.
Nevertheless, certain models assume this symmetry.
Experimentally, \eq{Go-sum-rule-definition} is
violated rather strongly \cite{Gottfried-exp}. Let us see what does the
chiral theory predict for the r.h.s. of \eq{Go-sum-rule-definition}.
Integrating \eq{triunpf} over x from $-1$ to 0 we find:

\beq
\int_0^1 dx [\bar u(x) - \bar d(x)]
=(2T_3) \frac{N_c i}{24\pi I}\int \frac{d \omega}{2 \pi}
\Sp \Bigl[ \tau^a (1+\gamma^0\gamma^3) \theta (-\omega-p^3 )
\frac1{\omega - H}\tau^a \frac1{\omega - H} \Bigr] \,.
\la{Go-sum-rule-expression}\eeq

It can be checked that this quantity vanishes in the leading order of the
gradient expansion, therefore the Gottfried sum rule is
satisfied only in the limit of very large solitons. For real
solitons the expression \ra{Go-sum-rule-expression} is generally
non-zero.

The Gottfried sum rule has been analyzed in the context of the chiral
quark-soliton model by Wakamatsu ~\cite{Wakamatsu-go}. Several suggestions
for the r.h.s. of the sum rule have been considered in that paper,
however none of them coincides literally with the exact result
\ur{Go-sum-rule-expression}.

\newpage

\section{Isovector polarized distribution}
\setcounter{equation}{0}

The polarized quark distribution function (see e.g.~\cite{Anselmino-PhRep})
is given by

\beq
\Delta q_f(x)= 2S_3 \frac{1}{4\pi}
\int\limits_{-\infty}^\infty dz^0  e^{ixM_Nz^0}
\bra{P,S} \psi_f^+(0) (1+\gamma^0\gamma^3)\gamma_5 \psi_f(z) \ket{P,S}
\Bigr|_{z^3=-z^0,\>z_\perp=0}\,,
\la{pol1}\eeq
where $S_3$ is the spin of the nucleon in its rest frame.
At negative $x$ the function $\Delta q_f(x)$ has the meaning of the
polarized antiquark distribution,

\begin{equation}
\Delta q_f(x) =\Delta \bar q_f(-x)
\la{Deltaminus}\end{equation}
(note the opposite sign as compared to the relation \ur{q-q-qbar} for the
unpolarized distributions!)

One can easily check that in the leading order of the $1/N_c$
expansion only the {\em isovector} polarized distribution
survives.  We remind the reader that in the case of unpolarized
distributions, on the contrary, the {\em singlet} distribution is large in
$N_c$.  Similarly to the derivation of \eq{distribution-occupied} from
\eq{main} we can write

\beqr
&&
\Delta u(x) - \Delta d(x)
= (2S_3)  \frac{N_c M_N}{2\pi}
\int\limits_{-\infty}^\infty dz^0  e^{ixM_Nz^0}
\int dR\; \phi_{T_3 S_3}^\dagger(R)
\nn
&&
\cdot
\sum\limits_{\scriptstyle n\atop \scriptstyle
{\rm occup.}}
e^{-iE_n z^0}
\bra n R^\dagger \tau^3 R
(1+\gamma^0\gamma^3)\gamma_5 \exp(-iz^0 p^3)
\ket n \phi_{T_3 S_3}(R)\,.
\la{tripol1}\eeqr

Computing the rotational matrix element with the rotational wave functions
$\phi(R)$ (see \eq{rotme}) we get:

\beqr
&&
\Delta u(x) - \Delta d(x) = - (2T_3)  \frac{N_c M_N}{6\pi}
\nn
&&
\cdot
\int\limits_{-\infty}^\infty dz^0  e^{ixM_Nz^0}
\sum\limits_{\scriptstyle n\atop \scriptstyle
{\rm occup.}}e^{-iE_n z^0}
\bra n \tau^3 (1+\gamma^0\gamma^3)\gamma_5 \exp(-iz^0 p^3)\ket n\,.
\la{h-L:nonsinglet:general}\eeqr
One has to take here $T^3=1/2$ for the distribution functions in a
proton and $T^3=-1/2$ for those in a neutron.
This sum over occupied states can be rewritten in terms of
the functional trace:

\begin{eqnarray}
&&
\Delta u(x) - \Delta d(x) = -\frac{1}{3} (2T_3)  N_c M_N
\nonumber\\
&&
\cdot
\int\limits_{-\infty}^{ E_{\rm lev}+0} d\omega\;
\mbox{Sp}\left[ \delta(\omega-H) \delta(\omega+p^3-xM_N)
\tau^3 (1+\gamma^0\gamma^3)\gamma_5
\right]
- (H\to H_0).
\end{eqnarray}
Integrating this equation over $x$ we reproduce the Bjorken
sum rule:

\begin{equation}
\int\limits_{0}^{1} dx [\Delta u(x) -\Delta d(x) + \Delta \bar u(x)
- \Delta \bar d(x)] = 2T_3\, g_A,
\la{g-A:sum-rule} \end{equation}
where $g_A$ is the nucleon axial constant. In deriving this sum rule
one has to keep in mind that the nucleon axial constant $g_A$ in the
leading order in $N_c$ is given by the following functional trace
\cite{W,MG}:

\beq
g_A =  - \frac{N_c}{3}
\int\limits_{-\infty}^{ E_{\rm lev}+0} d\omega\;
\mbox{Sp}\left[ \delta(\omega-H)
\tau^3 \gamma^0 \gamma^3 \gamma_5
\right]- (H\to H_0)\,.
\la{g-A}\eeq
It is understood that one has to calculate this trace with finite
pion mass and only then go to the chiral limit.

\section{Singlet polarized distribution}
\setcounter{equation}{0}

For this structure function one should replace $\tau^3$ in \eq{tripol1} by
a unity flavour matrix, hence $R^\dagger \tau^3 R$ is replaced by $1$. For
that reason the matrix element in \eq{tripol1} is zero in the lowest order
in the soliton rotation, and one has to expand the quark Green function to
the first order in the angular velocity, $R^\dagger\dot R$, as for the
isovector unpolarized distribution. Combining thus \eq{tripol1} and
\eq{triunpol1} we get

\begin{eqnarray} &&
\Delta u(x) + \Delta d(x) =
\frac{iM_N N_c}{2\pi} (2S_3)
\int\limits_{-\infty}^\infty dz^0  e^{ixM_Nz^0}
\int d^3{\bf X} \int dR \, \phi_{S_3T_3}^\dagger(R)
\nn
&&
\cdot
\Tr \Bigl((1+\gamma^0\gamma^3)\gamma_5\bra{z^0,{\bf z}-{\bf X}}
\frac{1}{i\partial_t - H}(- i R^\dagger \dot R )
\nn
&&
\cdot
\frac{1}{i\partial_t - H}\ket{0,-{\bf X}}  \Bigr)
\Bigr|_{z^3=-z^0,\>z_\perp=0}\phi_{S_3T_3}(R) \,.
\la{sinpol1}\end{eqnarray}

Making again the quantization substitution
$R^\dagger \dot R \rightarrow iS^a\tau^a/(2I)$ and integrating over the
soliton orientations $R$ we get a representation similar to
\eq{nonsinglet-distribution-result}:

\begin{eqnarray}
&&
\Delta u(x) + \Delta d(x) = \frac{iN_cM_N}{8\pi I}
\int\limits_{-\infty}^\infty dz^0  e^{ixM_Nz^0}\int d^3{\bf X}
\nonumber\\
&&
\cdot \Tr \Bigl( (1+\gamma^0\gamma^3)\gamma_5
\bra{z^0,{\bf z}-{\bf X}}\frac{1}{i\partial_t - H}\, \tau^3
\frac{1}{i\partial_t - H}\ket{0,-{\bf X}}\Bigr)
\Bigr|_{z^3=-z^0,\>z_\perp=0}\,.
\la{sinpol2}\end{eqnarray}
Performing the same steps as in subsection 4.2 we rewrite it through
the functional trace:

\beq
\Delta u(x)+\Delta d(x)
= \frac{iN_cM_N}{4 I} \int \frac{d\omega}{2\pi}
\Sp\Bigl( \gamma^0\gamma^3\gamma_5 \delta(\omega+p^3-xM_N)
\frac{1}{\omega - H}\, \tau^3 \frac{1}{\omega - H}\Bigr) \,,
\la{sinpol3}\eeq
which is similar in spirit to \eq{triunpf}. Again, repeating the
derivation of subsection 4.2 we can write the singlet polarized
distribution as a double sum over levels:

\begin{eqnarray}
&&
\Delta u(x) +\Delta d(x) = \frac{N_cM_N i}{8\pi I}
\sum\limits_{m,n} \int \frac{d^3{\bf p}}
{(2\pi)^3} \frac{\Phi_n^\dagger({\bf p})(1+\gamma^0\gamma^3)\gamma_5
\Phi_m({\bf p})}{(p^3 - xM_N - E_m)(p^3 - xM_N - E_n)}
\nonumber\\
&&
\cdot
\int \frac{d^3{\bf p^\prime}}{(2\pi)^3}
\Phi_n^\dagger({\bf p^\prime})\tau^3\Phi_m({\bf p^\prime})\,,
\end{eqnarray}
understood with the same $i\epsilon$ prescription as in \eq{ieps}.

Integrating \eq{sinpol3} over $x$ we obtain the fraction of the nucleon
spin carried by quarks:

\beq
\Delta u+\Delta d \equiv \int\limits_{-1}^{1} dx [\Delta u(x)+\Delta d(x)]
= \frac{iN_c}{4 I} \int \frac{d\omega}{2\pi}
\Sp \Bigl( \gamma^0\gamma^3\gamma_5
\frac{1}{\omega - H}\, \tau^3 \frac{1}{\omega - H}\Bigr)=g_A^{(0)} \,,
\la{sinpolsr}\eeq
which coincides, as it should, with the expression for the nucleon singlet
axial constant $g_A^{(0)}$ in the chiral quark-soliton model \cite{WY,BPG}.
The model calculation of this quantity gives
$g_A^{(0)}\approx 0.36$~\cite{BPG}; the rest of the nucleon spin (at low
$q^2$!) is carried by the orbital moment of the constituents and of the
distorted Dirac continuum~\cite{WY}. At higher values of $q^2$ an
increasing portion of the nucleon spin is carried by gluons.

\newpage

\section{Expressing distributions directly through the so\-li\-ton field}
\setcounter{equation}{0}

All expressions for the quark distributions we have derived above are in
fact certain functionals of the self-consistent pion field $U_c({\bf x})$
which binds the nucleon. It would be helpful to write down these
functionals in a more explicit form. That can be done but at a price
of additional approximations which, however, appear to be rather accurate
in practice.

A good place to start from is \eq{ImGF} which relates structure
functions to the imaginary part of the quark propagator in the background
field of the soliton, $U_c({\bf x})$. Expanding the propagator \ur{FGf}
in the derivatives of the pion field $\partial U_c$ up to the second order
we obtain two contributions to the singlet distribution:

\beqr &&
\sum_f q_f(x)\approx \frac{N_cM_NM^2}{2\pi}\mbox{ Im }
\int\frac{d^3{\bf k}}{(2\pi)^3}\int\frac{d^4 p}{(2\pi)^4}
\frac{2\pi\delta(p^0+p^3-xM_N)}
{(M^2-(p-k)^2-i0)(M^2-p^2-i0)^2}
\nn
&&
\cdot\Bigl\{(M^2-p^2)\mbox{ Tr}\left[\tilde U^{-\gamma_5}(-{\bf k})
\Dirac{k}\tilde U^{-\gamma_5}({\bf k})(\gamma^0+\gamma^3)\right]
\nn
&&
+\mbox{ Tr}\left[\Dirac{p}\Dirac{k}\tilde U^{-\gamma_5}(-{\bf k})
\Dirac{k}\tilde U^{-\gamma_5}({\bf k})(\gamma^0+\gamma^3)\right]\Bigr\},
\la{int1}\eeqr
where we have introduced the Fourier transform of the pion field,

\beq
{\tilde U}({\bf k}) = \int d^3 {\bf r}\; e^{-i({\bf k}{\bf r})}\;
[U_c({\bf r})-1].
\la{tilde-U-q}\eeq

The integration over the quark loop momenta $p$ can be easily performed:
one first uses the $\delta$-function to integrate over
$p^0$, then one integrates over $p^3$ taking the residues
of one of the denominators. The condition that the poles in $p^3$ lie on
different sides of the integration axis so that one gets a non-zero
imaginary part, is $k^3>xM_N$ for $x>0$ and $k^3<xM_N$ for $x<0$. For
$x<0$ we change the dumb variable ${\bf k}\rightarrow -{\bf k}$, so that
the condition can be written in a common form as $k^3>|x|M_N$.
We get

\beqr &&
\sum_f q_f(x) \approx \mbox{sign}(x)\frac{N_cM_NM^2}{\pi}
\int\frac{d^3{\bf k}}{(2\pi)^3}\mbox { Tr }\left\{\tilde U({\bf k})
[\tilde U({\bf k})]^\dagger\right\}\theta(k^3-|x|M_N)
\nn
&&
\cdot\int\frac{d^2{\bf p}_\perp}{(2\pi)^2}\frac{M^2+{\bf p}_\perp^2}
{(M^2+{\bf p}_\perp^2+\kappa^2)^2},\;\;\;\;\;\;\;\;\;\;
\kappa^2=\frac{|x|M_N(k^3-|x|M_N){\bf k}^2}{(k^3)^2}\;>\;0.
\la{int2}\eeqr
The last integral is logarithmically divergent. We regularize it by the
Pauli--Villars method: one has to compute the structure functions replacing
the quark mass $M \rightarrow M_{PV}$ where $M_{PV}$ is the Pauli--Villars
regulator mass, multiply the obtained result by $M^2/M_{PV}^2$ and subtract
it from the original distribution. At the moment it is the only practical
way of regularizing the structure functions we know of, which does not
violate casuality and analyticity, see above. Note that in calculating
static characteristics of the nucleon the requirements on the
regularization method are not so restrictive. The value of the
Pauli--Villars mass is fixed from the value of the $F_\pi$ constant
\cite{DPP}:

\beq
F_\pi^2 = 4N_c\int\frac{d^4k}{(2\pi)^4}\;\frac{M^2}{(M^2+k^2)^2}
-4N_c\frac{M^2}{M_{PV}^2}\int\frac{d^4k}{(2\pi)^4}\;
\frac{M_{PV}^2}{(M_{PV}^2+k^2)^2}=\frac{N_cM^2}{4\pi^2}
\ln\frac{M_{PV}^2}{M^2}.
\la{Fpi}\eeq

Integrating \eq{int2} over ${\bf p}_\perp$ and performing the Pauli-Villars
regularization we obtain finally for the singlet structure function:

\beqr
&&
\sum_f q_f(x)\approx \mbox{sign}(x)\frac{N_cM_NM^2}{4\pi^2}
\int\frac{d^3{\bf k}}{(2\pi)^3}\theta(k^3-|x|M_N)
\mbox{ Tr }\left(\tilde U({\bf k})[\tilde U({\bf k})]^\dagger\right)
\nn
&&
\cdot\left[\ln\frac{M_{PV}^2+\kappa^2}{M^2+\kappa^2}-
\frac{\kappa^2(M_{PV}^2-M^2)}{(M_{PV}^2+\kappa^2)(M^2+\kappa^2)}\right].
\la{interpol}\eeqr

We call it {\em interpolation formula} as \eq{interpol} becomes exact in
three limiting cases: i) low momenta, $|\partial U|\ll M$, ii) high
momenta, $|\partial U|\gg M$, iii) any momenta but small pion fields,
$|\log U| \ll 1$. Therefore, we expect that the interpolation formula
has a good accuracy also in a general case. As compared to exact
calculations involving summation over all levels the use of the
interpolation formula gives an enormous gain in computing time: for a given
profile of the pion field in the nucleon one has to compute numerically
just three integrals.

If the spatial size of the soliton is large, meaning that momenta
${\bf k}$ in \eq{interpol} are small, one can neglect $\kappa^2$
in the quark loop integral and get a simple formula:

\beq
u(x)+d(x) \approx \mbox{sign}(x) F_\pi^2 M_N \int
\frac{d^3 {\bf k}}{(2\pi)^3} \theta(k^3 - M_N|x|) \>
\Tr\left(\tilde U({\bf k}) [\tilde U({\bf k})]^\dagger \right)
\la{gradient2}\eeq

This is a remarkable equation. First, it shows that if the nucleon
has the size $r_0$ the singlet structure function is concentrated
around the values of $x \sim 1/(r_0M_N)$. Second, we notice that
\eq{gradient2} is {\em non-analytic} in the pion field momenta.
It means that one cannot make a gradient expansion either for the
structure functions or for their moments. This situation is new:
all static characteristics of the nucleon admit expansion in the
gradients of the pion field \cite{DPP}. By the way, it means that
there is no reasonable way to extract structure functions from the
Skyrme model (which contains just two and four derivatives of the pion
field), even if one invents a way to identify leading-twist operators
in that model.

It is amusing that the second moment of the distribution \ur{gradient2}
is $2/3$ instead of unity. Indeed, integrating \eq{gradient2} over $x$
with $x$ one gets for the second moment of the singlet structure
function

\beq
M_2\approx \frac{F_\pi^2}{M_N}\int\limits_{k^3>0}
\frac{d^3{\bf k}}{(2\pi)^3}\;k_3^2\;
\Tr\left(\tilde U({\bf k}) [\tilde U({\bf k})]^\dagger \right)
\la{M2}\eeq
where $k_3^2$ can be replaced by ${\bf k}^2/3$ for the spherically
symmetric hedgehog field. At the same time the leading contribution to
the large-size nucleon mass is given just by the kinetic-energy term
of the chiral lagrangian,

\beq
M_N\approx \frac{F_\pi^2}{4}\int d^3{\bf x}\mbox { Tr }
[\partial_i U({\bf x})\partial_i U^\dagger({\bf x})].
\la{kinterm}\eeq
Comparing \eqs{M2}{kinterm} we see that in the large-size approximation
the nucleon energy at rest computed from the total energy carried by its
constituents is $2/3$ that of its mass! Exactly this ``$2/3$" paradox
has been discovered some time ago in a similar situation by Lorentz
\cite{Lor} who attempted to calculate the energy of the electron from a
charge distribution bound by some unknown forces of non-electromagnetic
origin. He found an interesting relation: $E=\frac{2}{3}mc^2$. The paradox
is resolved when one exploits the equation of motion for constituents. In
our case \eq{kinterm} has no minimum, however the full functional for the
nucleon mass, including the discrete level, has. As shown in subsection~3.5
the use of the equation of motion for the pion field is essential in
establishing the correct momentum sum rule.

\Eq{gradient2} shows that in the leading order of the above expansion
the singlet quark distribution coincides with the antiquark one. In order
to obtain a nonvanishing result for the difference of singlet quark and
antiquark distributions one has to go to the next (third) order in
expanding the quark propagator. This time we perform the expansion
in the coordinate space and get an elegant expression:

\beqr
&&
\sum\limits_f[q_f(x)-\bar q_f(x)]
\approx \frac{N_cM_N }{4\pi^3}\int d^3 {\bf y} \int_{-\infty}^{\infty}
d\xi\; e^{-i\xi M_N x}  \int\limits_0^1 d\alpha \int\limits_0^\alpha d\beta
\int\limits_0^\beta d\gamma
\nn
&&
\cdot \epsilon_{ijk} \Tr\left(
[\partial_i U({\bf y}+\alpha\xi {\bf e}_3)]
[\partial_j U({\bf y}+\beta\xi {\bf e}_3)]^\dagger
[\partial_k U({\bf y}+\gamma\xi {\bf e}_3)]
[U({\bf y})]^\dagger\right).
\la{q-minus-qbar-gradient}\eeqr

It is remarkable that this result is consistent with the baryon number sum
rule \ra{baryon-number:sum-rule}. Indeed, for the solitons of a large
size (for which the above expansion is justified) the baryon
number coincides \cite{DPP} with the winding number of the pion field:

\bq{
B =  \frac{1}{24\pi^2} \int d^3 y \epsilon_{ijk}
\Tr\left[ (U^\dagger\partial_i U)
(U^\dagger\partial_j U)
(U^\dagger\partial_k U)
\right].
\la{B-winding-number}
}
Integrating \eq{q-minus-qbar-gradient} over $x$ one immediately obtains

\bq{
\int\limits_{0}^1 dx
\sum\limits_f[q_f(x)-\bar q_f(x)]
= N_c B
}
where $B$ is given by the soliton winding number,
\eq{B-winding-number}.

Turning to the polarized distributions we remind the reader that in the
leading order of the $1/N_c$ expansion only the isovector function
$\Delta u(x) - \Delta d(x)$ survives. Similarly to \eq{interpol} we obtain:

\beqr
&&
\Delta u(x) - \Delta d(x)
\approx\frac{1}{3}\;(2T_3)\frac{N_cM_NM^2}{4\pi^2}
\int\frac{d^3{\bf k}}{(2\pi)^3}\theta(k^3-|x|M_N)
\mbox{ Tr }\left(\tilde U({\bf k})[\tilde U({\bf k})]^\dagger\tau^3\right)
\nn
&&
\cdot
\left[\ln\frac{M_{PV}^2+\kappa^2}{M^2+\kappa^2}-
\frac{\kappa^2(M_{PV}^2-M^2)}{(M_{PV}^2+\kappa^2)(M^2+\kappa^2)}\right],
\la{polinterpol}\eeqr
where $\kappa^2$ is given in \eq{int2}. For large-size solitons it can be
simplified to

\beq
\Delta u(x) - \Delta d(x)
\approx \frac{1}{3}\, (2T_3) F_\pi^2 M_N \int \frac{d^3
k}{(2\pi)^3}\, \theta(k^3 - M_N|x|) \Tr \left(U({\bf k}) [U({\bf
k})]^\dagger\tau^3 \right).
\la{polgrad}\eeq
Note that \eqs{polinterpol}{polgrad} are even in $x$; the odd part of the
Dirac continuum contribution to $\Delta u(x) - \Delta d(x)$ arises in the
next order in the interpolation formula and is thus small.

Equations \urss{interpol}{q-minus-qbar-gradient}{polinterpol}
can be used to make a quick estimate of the structure functions, before
one plugs into exact numerics which, in any case, is rather laborious.
One should not forget, however, to add the contribution of the discrete
level. To be consistent, a contribution of the discrete level with a
Pauli--Villars mass to a distribution function $f(x)$ should be subtracted:

\beq
f^{\rm lev}(x)\rightarrow f_M^{\rm lev}(x)-\frac{M^2}{M_{PV}^2}
f_{M_{PV}}^{\rm lev}(x).
\eeq

 \section{Numerical results and discussion}
\setcounter{equation}{0}

We have calculated numerically two structure functions surviving in the
leading order in $N_c$: the singlet unpolarized distribution (section 3)
and the isovector polarized one (section 5); in both cases we get quark
and antiquark distributions separately.

A variational estimate of the best profile of the pion-field soliton
(see \eq{hedge}) has been performed in ref. \cite{DPP} yielding
for $M=350$~MeV

\beq
P(r)=-2\;\mbox{arctg}\left(\frac{r_0^2}{r^2}\right),\;\;\;\;\;\;
r_0\approx 1.0/M,\;\;\;\;\;\;M_N\approx 1170\;MeV.
\la{varprof}\eeq
This profile function has a correct behaviour at small and large distances
and is stable in respect to small perturbations. In our numerics we use
the analytical profile~\ur{varprof}: its difference with the exact solution
is small.  The value of the Pauli-Villars mass needed for the
regularization is found from \eq{Fpi} to be $M_{PV}\approx 560\;MeV$.

We estimate the Dirac continuum contribution using the
interpolation formulae: \eq{interpol} for the unpolarized singlet
distribution and \eq{polinterpol} for the isovector polarized
distribution. In fact we have performed the exact computation of the Dirac
continuum contribution by direct summation over all levels: as illustrated
by Figs.1,2, its deviation from the interpolation formula proves to be
small.  The description of exact computations will be a subject of a
separate publication \cite{DPPPW}. Since the reader may wish to repeat the
calculations with his or her favourite set of parameters, we suggest
the use of the interpolation formulae which give reliable
approximations to the structure functions but can be computed in a few
minutes on a PC.

As stressed many times in the paper, the quark distribution $q(x)$ is a
single function defined for both positive and negative $x$'s. At $x<0$
it gives actually the distribution of antiquarks, $\bar q(x) = - q(-x)$.
The $x$-even combination, that is $q(x)-\bar q(x),\;x>0$,
 which is actually the baryon number density,
 is finite
when one takes the ultraviolet cutoff to infinity. Moreover, it {\em should
not} be regularized at all as it corresponds to the imaginary part of the
effective chiral action \ur{FI}. Probably, an ultraviolet regularization
can be introduced which takes into account automatically the
non-renormalization of the imaginary part of the chiral action. The
original regularization by the momentum-dependent constituent quark mass
$M(p)$ \cite{DP} seems to satisfy this requirement \cite{BR}. In this
paper, however, we mimic that regularization {\em i}) by the relativistic
Pauli--Villars regularization for the logarithmically divergent $x$-odd
part of $q(x)$, that is for $q(x)+\bar q(x)$, {\em ii}) by making
no regularization for the $q(x)-\bar q(x)$ distribution.
In the case of isovector polarized distributions, the logarithmically
divergent combination is $[\Delta u(x)-\Delta d(x)]
+[\Delta \bar u(x)- \Delta\bar d(x)],\;x>0$, corresponding to the
$g_1^p-g_1^n$ structure function; the other combination originates from
the imaginary part of the effective chiral action and should not be
regularized.

Figures 1--5 present our results for the following distribution functions:
\begin{itemize}
\item  Fig.1: $\;x[u(x)+d(x)+\bar u(x)+\bar d(x)]/2$;
\item  Fig.2: $\;x[u(x)+d(x)-\bar u(x)-\bar d(x)]/2$;
\item  Fig.3: $\;x[\bar u(x)+\bar d(x)]/2$;
\item  Fig.4: $\;x[\Delta u(x)-\Delta d(x)
+\Delta\bar u(x)-\Delta\bar d(x)]/2$;
\item  Fig.5: $\;x[\Delta\bar u(x)-\Delta\bar d(x)]/2$.
\end{itemize}
In Figs. 1 and 4 we plot the contributions of the discrete level
and that of the Dirac continuum (computed via the interpolation formula) as
well as their sum separately. The exact calculation of the sum is also
shown in  Fig.1. It can be seen that the interpolation formula gives a
good approximation to the exact calculations. The same is seen from Fig.2.
The discrete-level contributions are given in Appendix B.
One can see that our quark and antiquark distribution functions are
positive (Fig.3), and that the contribution of the Dirac continuum is
significant.

It should be emphasized that the distribution of {\em antiquarks} arising
from the discrete level (see \eq{valence1}) is definitely negative
(and sizeable!) \footnote{In the extreme case of a very strongly
bound discrete level when it approaches the lower continuum, this
level would not produce quarks at all -- only antiquarks,
but with a negative sign!} Positivity of the structure functions is
restored only when one includes the contribution of the Dirac continuum.
Indeed, the existence of a non-trivial discrete level is due to a strong
mean pion field in the large-$N_c$ nucleon.  Taking into account only the
discrete-level contribution to the structure functions means ignoring the
degrees of freedom stored in the field creating that level, hence the
"negative probability", $\bar q(x)<0$.  Adding the Dirac continuum
contribution makes the system complete, and all the probabilities become
positive.  We have demonstrated above that all the general sum rules are
fulfilled only when one adds the discrete level and the Dirac continuum
contributions together.

There seems to be a lesson here for all variants of bag models. It is
usually assumed that the three quarks in the bag give rise only to quark
distributions, however they inevitably produce also {\em negative}
antiquark distributions. The baryon number computed from the quark
distributions only, turns out to be {\em less} than unity; to restore unity
one has to subtract the negative distribution of the antiquarks. This
circumstance is not often emphasized. To cure this decease one would need
to add the structure function arising from the forces which keep the three
quarks bound, in this case from the bag surface.

To stress it once again: the discrete level produces not only valence
quarks (in the sense used in the deep inelastic scattering phenomenology),
whereas the Dirac continuum produces non-equal distributions of
quarks and antiquarks (see Fig.3), therefore its contribution should not be
identified with the quark-antiquark sea of the DIS folklore.

In the large $N_c$ limit the nucleon is heavy, so we have neglected its
recoil. For that reason the structure functions do not automatically go to
zero at $x=1$. However at $x\gg 1/N_c$ the distributions behave as
$\sim\exp(-\mbox{const}\cdot N_cx)$; numerically, even for $N_c=3$ all
distributions computed are very small at $x\approx 1$. Nevertheless, we
should caution those who may wish to reconstruct the structure
functions from the moments: at $n\geq N_c$ the $n$-th moments become
sensitive to the tail of the structure functions at $x\sim 1$ where
$1/N_c$ corrections become 100\% important, and the moments become thus
absolutely unreliable.

We remind the reader that our calculations refer to the leading-twist
distribution functions at a normalization point of about $600$~MeV. In
order to make a meaningful comparison with the data one has to use our
distributions as initial conditions for the standard perturbative evolution
of the structure functions to higher values of $q^2$ where the actual data
are available.  This evolution takes into account the bremsstrahlung of
gluons and also their conversion into quark-antiquark pairs. It is well
known that the perturbative evolution makes the distributions more ``soft".
Therefore, it can well wash out the quark excess at high $x$ and cure the
deficit at low $x$. This part of the investigation remains to be done.

However, we can still make a comparison if not directly with the data,
then with the parametrization of the data at a low normalization point
$\mu\approx 600\;MeV$ performed recently by Gl\"uck, Reya {\em et al.}
\cite{GRV,GR2}. The standard perturbative evolution of their distributions
describes well all the existing data at larger values of $q^2$. It is
known, though, that the evolution in the opposite direction is
highly unstable.  However, we believe that the low-point distributions
suggested in refs. \cite{GRV,GR2} are reasonable, and we compare our curves
to their in Figs. 1--4.

Despite using the parametrization of the data at a low value of $q^2$,
the ``experimental" distributions appear to be more ``soft" than the
calculated ones. On the whole it looks as if we have determined the
distributions at an even lower normalization point than that of refs.
\cite{GRV,GR2}. Since $\alpha_s$ at these momenta are large it may require
quite a short evolution range to move our distributions to those of
\cite{GRV,GR2}.

The calculated isovector polarized distribution shown in Fig.3 appears
to be systematically less than the parametrization of ref. \cite{GR2}.
We would like to make three comments on this discrepancy. First, we think
that the parametrization of the polarized distributions is less reliable
than that of the unpolarized ones as it is based on less data with larger
errors. In particular, the authors of \cite{GR2} have assumed that
$\Delta \bar u(x) - \Delta \bar d(x)=0$ which is not confirmed in the model
we are considering: this quantity appears to be of the same order as
$\Delta \bar u(x) + \Delta \bar d(x)$ of ref. \cite{GR2}, see
Fig.5. Second, by choosing the parameters of the soliton in \eq{varprof} we
have, unfortunately, implanted a somewhat small value of the $g_A$ constant
to which the distribution of Fig. 4 is normalized according to the Bjorken
sum rule: it is $g_A\approx 0.96$ instead of $1.25$. Third, it is known
\cite{WY,Chr} that this channel is particularly sensitive to the $1/N_c$
corrections which are altogether neglected in this paper.

On the whole we get a reasonable description of several parton
distributions without adjusting the parameters of the model to make a
``best fit".

Finally, we would like to comment on a recent work \cite{WGR} where
quark distributions have been estimated in the Nambu--Jona-Lasinio model
which, after certain simplifications, is reduced to the chiral
quark-soliton model of ref.\cite{DPP} considered here. Only the
contribution of the discrete level
to the unpolarized structure functions has been studied in
that work. As explained above, this
approximation leads to a number of inconsistencies. For example, one
obtains a wrong (negative) sign for the singlet antiquark distribution
function, while the baryon number obtained from integrating the quark
distribution only, is less than unity. As to the isovector distributions
also considered in \cite{WGR}, their basic eq. (25) does not seem to agree
with our result given by \eq{isofin}.

\section{Conclusions}
\setcounter{equation}{0}

At large number of colours the nucleon can be viewed as a heavy
semiclassical body whose $N_c$ ``valence" quarks are bound by a
self-consistent pion field. The energy of the pion field is given
by the effective chiral lagrangian and coincides with the aggregate
energy of the Dirac sea of quarks (the free continuum subtracted).
Therefore, to compute the deep-inelastic structure functions in
the large $N_c$ limit, it is sufficient to calculate the quark and
antiquark distributions arising from the discrete level occupied by
quarks, and from the (distorted) negative-energy Dirac continuum.
In contrast to all variants of the bag model, the completeness of the
states involved guarantees the consistency of the calculations.
Indeed, we have checked the validity of the baryon number,
isospin, total momentum and of the Bjorken sum rules. We have
also derived an expression for the r.h.s. of the Gottfried
sum rule, which differs from the previously suggested ones.
To our knowledge, it is the first time that the nucleon structure
functions are theoretically calculated in a relativistic model which
preserves all general properties of parton distributions.

In the academic limit of a very weak mean pion field the Dirac continuum
reduces to the free one (and should be subtracted to zero) while the
discrete level joins the upper Dirac continuum. In that limit there
are no antiquarks, while the distribution function of quarks becomes
$\delta(x-1/N_c)$. In reality there is a non-trivial mean pion
field which {\em i}) creates a discrete level,
{\em ii}) distorts the negative-energy Dirac continuum. As a result, the
above $\delta$-function is smeared significantly (but still in the range of
the order of $1/N_c$), and non-zero antiquark distributions appear even at
a low normalization point. It should be stressed that antiquarks come not
only from the Dirac continuum but also from the discrete level, whereas
the Dirac continuum produces nonequal distributions of quarks and
antiquarks.

As to the gluon distribution, it depends on the details of the
ultraviolet regularization of the effective chiral theory (corresponding to
the ``formfactor" of the constituent quark), and for that reason we have not
attempted to determine it here. Also, we have not tried to make a ``best
fit" to the parametrizations of the data \cite{GRV,GR2} by adjusting the
parameters of the model, such as the constituent quark mass and the way one
regularizes the effective chiral theory. In fact we have used practically
the same set of parameters as deduced \cite{DP,DPP} from
instantons. It is remarkable that we are able to reproduce the basic
features of the distributions, despite neglecting completely the $1/N_c$
corrections. It would  certainly be preferable to use our leading-twist
distributions as initial conditions to the standard perturbative evolution
to higher values of $q^2$ where a direct comparison to the data is
possible, instead of comparing to the parametrizations of the data at a low
normalization point, which to some extent is model-dependent.

We have shown that, from the point of view of large $N_c$,
all distributions can be divided into ``large" and ``small" ones.
The large ones are the singlet unpolarized quark and antiquark
distributions and the isovector polarized distributions. The
isovector unpolarized and the singlet polarized are, parametrically, $N_c$
times smaller, which seems to be confirmed experimentally despite that in
reality $N_c$ is only three.

We have found that the structure functions are non-analytic
in the pion field momenta, though one can still write exact
expressions for the structure functions in the limit of large-size
nucleons, see section~7. For arbitrary sizes we have derived
{\em interpolation formulae} for the structure functions which allow
one to compute the structure functions in a few minutes on a PC and
reproduce the results of exact calculations to  good accuracy.

We have restricted ourselves to the case of $u$ and $d$ quarks only,
though the generalization to three flavours is quite simple in the
collective-quantization technique -- see, e.g., ref. \cite{B}.

The methods developed in this paper can be easily generalized to
higher-twist observables, like the $g_2,h_{T}...$ structure functions.
They can be also used to estimate power corrections to the numerous
structure functions at relatively low $q^2$.

\vskip 1.5true cm

{\large\bf Acknowledgements}
\vskip .5true cm
This work has been supported in part by the NATO Scientific Exchange grant
OIUR.LG 951035, by the Deutsche Forschungsgemeinschaft and by
COSY (J\"ulich). The Russian
participants acknowledge the hospitality of Bochum University; C.W.
acknowledges the hospitality of the Petersburg Nuclear Physics Institute,
where parts of the work has been done. P.P. is being supported by the
A.v.Humboldt Foundation. D.D. is grateful to J.~Bjorken and A.~Mueller for
an inspiring conversation. It is a pleasure to thank Georges Ripka for
useful discussions at the initial stage of the work, and Klaus Goeke for
encouragement and multiple help.

\newpage

\appendix
\renewcommand{\theequation}{\Alph{section}.\arabic{equation}}

\section{Expressing distributions through the Feyn\-man pro\-pa\-ga\-tor}
\setcounter{equation}{0}

The Feynman Green function in a stationary background field is defined as

\beqr
&&
G_{ij}(x,y)=i\bra 0\mbox{T} \left\{\psi_i(y) \bar\psi_j(x) \right\}\ket 0
\nn &&
=i\theta(x^0-y^0) \sum\limits_{\scriptstyle n\atop
\scriptstyle {\rm non-occup.}} \exp[-i E_n(x^0-y^0)] \Phi_{n,i}({\bf x})
\bar\Phi_{n,j}({\bf y})
\nn &&
-i\theta(y^0-x^0)\sum\limits_{\scriptstyle n\atop
\scriptstyle {\rm occup.}} \exp[-iE_n(x^0-y^0)] \Phi_{n,i}({\bf x})
\bar\Phi_{n,j}({\bf y})\,.
\la{FGfdef}\eeqr
In the free case (no background field) it comes to

\beq
G_{ij}(x,y)=\int\frac{d^4p}{(2\pi)^4}\;e^{-i(p\cdot(x-y))}\;
\frac{M+\Dirac p}{M^2-p^2-i0}.
\la{free}\eeq
The sign convention is such that the imaginary part of the Feynman
propagator is positive.

Let us rewrite the singlet structure function, say, in representation
of \eq{singl1} through the Feynman propagator. To that end we divide
the full range of integration in $z^0$ in \eq{singl1} in two parts:
from 0 to $+\infty$ and from $-\infty$ to 0. Let us call these two
contributions to the full structure function $q_1(x)$ and $q_2(x)$
respectively. In the $q_2(x)$ part let us change the integration
variable $z^0\rightarrow -z^0$, so that the new variable also runs
from 0 to $+\infty$. We have from \eq{singl1}:

\beq
q_1(x)=\frac{N_cM_N}{2\pi}\int d^3{\bf X}\int_0^\infty e^{ixM_Nz^0}
\sum\limits_{\scriptstyle n\atop \scriptstyle {\rm occup.}} e^{-i E_n z^0}
\bar\Phi_n(-{\bf X})(\gamma^0+\gamma^3)\Phi_n(-z^0{\bf n}_3-{\bf X})\,,
\la{q1}\eeq

\beq
q_2(x)=\frac{N_cM_N}{2\pi}\int d^3{\bf X}\int_0^\infty e^{-ixM_Nz^0}
\sum\limits_{\scriptstyle n\atop \scriptstyle {\rm occup.}} e^{i E_n z^0}
\bar\Phi_n(-{\bf X})(\gamma^0+\gamma^3)\Phi_n(z^0{\bf n}_3-{\bf X})\,.
\la{q2}\eeq

Comparing \eq{q2} with the definition of the Feynman Green function
\ur{FGfdef} we see that it can be written as

\beq
q_2(x)=i\frac{N_cM_N}{2\pi}\int d^3{\bf X}\int_0^\infty e^{-ixM_Nz^0}
\mbox { Tr }G(z^0{\bf n}_3-{\bf X}, -z^0;\;-{\bf X},0)(\gamma^0+\gamma^3).
\la{q2G}\eeq

The other part, \eq{q1}, does not directly fit into the definition of the
Feynman Green function, however its complex conjugate does. Indeed, making
the complex conjugation of \eq{q1} and changing the integration variable
${\bf X}\rightarrow {\bf X}-z^0{\bf n}_3$ we get exactly the r.h.s. of
\eq{q2}, therefore $q_1^*(x)=q_2(x)$. Using \eq{q2G} we get for the
full singlet structure function

\beq
\sum_f q_f(x)=-2\mbox{ Im }\frac{N_cM_N}{2\pi}\int\! d^3{\bf X}
\int_0^\infty\! e^{-ixM_Nz^0}
\mbox { Tr }G(z^0{\bf n}_3-{\bf X}, -z^0;\;-{\bf X},0)(\gamma^0+\gamma^3).
\la{Im1}\eeq

Let us now pass to the Fourier transform of the propagator. For the
time-independ\-ent background field one writes:

\beq
G(x,y)=\int\frac{d^4p_1}{(2\pi)^4}\int\frac{d^4p_2}{(2\pi)^4}\;
2\pi\delta(p_1^0-p_2^0)S(p_1^0, {\bf p}_1,{\bf p}_2)e^{i(p_1\cdot x)
-i(p_2\cdot y)}.
\la{FtGf}\eeq
Putting it in \eq{Im1} we integrate over ${\bf X}$, which yields
$(2\pi)^3\delta({\bf p}_1-{\bf p}_2)$. Finally, we integrate \eq{Im1} in
$z^0$ and obtain

\beq
\sum_f q_f(x)=2\mbox{ Im }\frac{N_cM_N}{2\pi}\int \frac{d^4 p}{(2\pi)^4i}
\frac{\mbox{Tr}[S(p^0,{\bf p},{\bf p})(\gamma^0+\gamma^3)]}
{p^0+p^3-xM_N+i0}.
\la{Im2}\eeq

We have made this derivation starting from \eq{singl1} where the summation
over occupied levels is used. We could as well start from the equivalent
\eq{distribution-non-occupied} where summation goes over non-occupied
levels. Repeating the same steps as above we arrive to \eq{Im2} but with
the opposite overall sign and the opposite sign of $i0$ in the denominator.
Since the two formulas must be equivalent in a ``good" renormalization
scheme, it means that a non-zero imaginary part of the whole expression
arises solely from the $i\pi\delta(p^0+p^3-xM_N)$ pieces of the
denominators in both cases. Thus, we obtain:

\beq
\sum_f q_f(x)=-\mbox{ Im }\frac{N_cM_N}{2\pi}\int \frac{d^4 p}{(2\pi)^4}\;
2\pi\delta(p^0+p^3-xM_N)\mbox{ Tr }
[S(p^0,{\bf p},{\bf p})(\gamma^0+\gamma^3)],
\la{Im3}\eeq
which is \eq{ImGF} of the main text.

\section{Bound-state level}
\setcounter{equation}{0}

We present here the contributions of the discrete bound-state
level to the singlet unpolarized and to the isovector
polarized structure functions. This is the cases where the
discrete-level  contribution is well-defined and in fact large. The other
two structure functions considered in the paper are expressed through a
{\em double} sum over levels, hence the contribution of the discrete
level is not specific.

The bound-state level occurs in the grand spin $K=0$ and parity
$\Pi=+$ sector of the Dirac hamiltonian \ur{hU}. In that sector
the eigenvalue equation takes the form~\footnote{We change the sign
of the $\gamma_5$ matrix and hence of the profile function $P(r)$
as compared to ref. \cite{DPP}. The $\gamma_5$ matrix is now that
of Bjorken and Drell. The profile function is equal to $-\pi$ at
the origin.}:

\beq
\left(\begin{array}{cc}
 M \cos P(r) & -\frac{\partial}{\partial r}-\frac{2}{r} - M \sin P(r)\\
\frac{\partial}{\partial r} - M \sin P(r) & - M \cos P(r)
\end{array}\right)
\left(\begin{array}{c}
h_0(r) \\ j_1(r)
\end{array}\right)
=E_{\rm lev}
\left(\begin{array}{c}
h_0(r) \\ j_1(r)
\end{array}\right).
\la{H-K-0:Pi-plus}\eeq
We assume that the radial wave functions are normalized by the
condition

\beq
\int\limits_{0}^{\infty} dr \, r^2  [h_0^2(r) + j_1^2(r)] =1.
\eeq
We introduce the Fourier transforms of the radial wave functions,

\beq
h(k) = \int\limits_{0}^{\infty} dr \, r^2\, h_0(r) R_{k0}(r),\;\;\;\;\;\;
j(k) = \int\limits_{0}^{\infty} dr \, r^2\, j_1(r) R_{k1}(r),
\eeq
where
\beq
R_{kl}(r) =  \sqrt{ \frac{k}{r}} J_{l+\frac12}(kr)
= (-1)^l\sqrt{\frac{2}{\pi}} \frac{r^l}{k^l}
\left( \frac{1 }{r } \frac{d}{dr } \right)^l \frac{\sin kr }{r}.
\eeq

The bound-state level contribution to the singlet unpolarized
structure function can be simply obtained from the general
\eq{distribution-occupied}. We get:

\beq
[u(x)+d(x)]_{\rm val}(x) = N_c M_N \int\limits_{|xM_N-E_{\rm lev}|}^{\infty}
\frac{dk}{2k} \Biggl\{ h^2(k)
+ j^2(k)- 2 \frac{xM_N-E_{\rm lev}}{k} h(k) j(k)\Biggr\}.
\la{valence1}
\eeq
Note that the r.h.s. is positive for all values of $x$, in particular
at $x<0$ where \eq{valence1} determines in fact the antiquark distribution.
Since $\bar q(x)=-q(-x)$, it means that \eq{valence1} gives a
{\em negative} distribution of antiquarks at $x>0$. At the same time
it is easy to check by integrating \eq{valence1} that the baryon number sum
rule is fully saturated by the discrete-level contribution only. It means
that without the subtraction of the {\em negative} antiquark distribution
the baryon number is not conserved. Simultaneously it means that the total
baryon number of the Dirac continuum is zero, though locally in $x$ the
antiquark distribution from the Dirac continuum does not necessarily
coincide with the quark one, see the dotted line in Fig.2.

The bound-state contribution to the polarized isovector distribution
function in the proton is obtained from \eq{h-L:nonsinglet:general}:

\beqr
&&
\left[\Delta u(x) - \Delta d(x)\right]_{\rm val} = \frac13  N_c M_N
\!\!\int\limits_{|xM_N - E_{\rm lev}|}^{\infty}\!\!
\frac{dk}{2k} \Biggl\{ h^2(k) +\left[2 \frac{(xM_N - E_{\rm lev})^2}{k^2}
- 1 \right] j^2(k)
\nn
&&
- 2 \frac{(xM_N - E_{\rm lev})}{k} h(k) j(k)\Biggr\}.
\la{valence2}\eeqr

\newpage
\begin{figure}
 \vspace{-1cm}
\epsfxsize=16cm
\epsfysize=15cm
\centerline{\epsffile{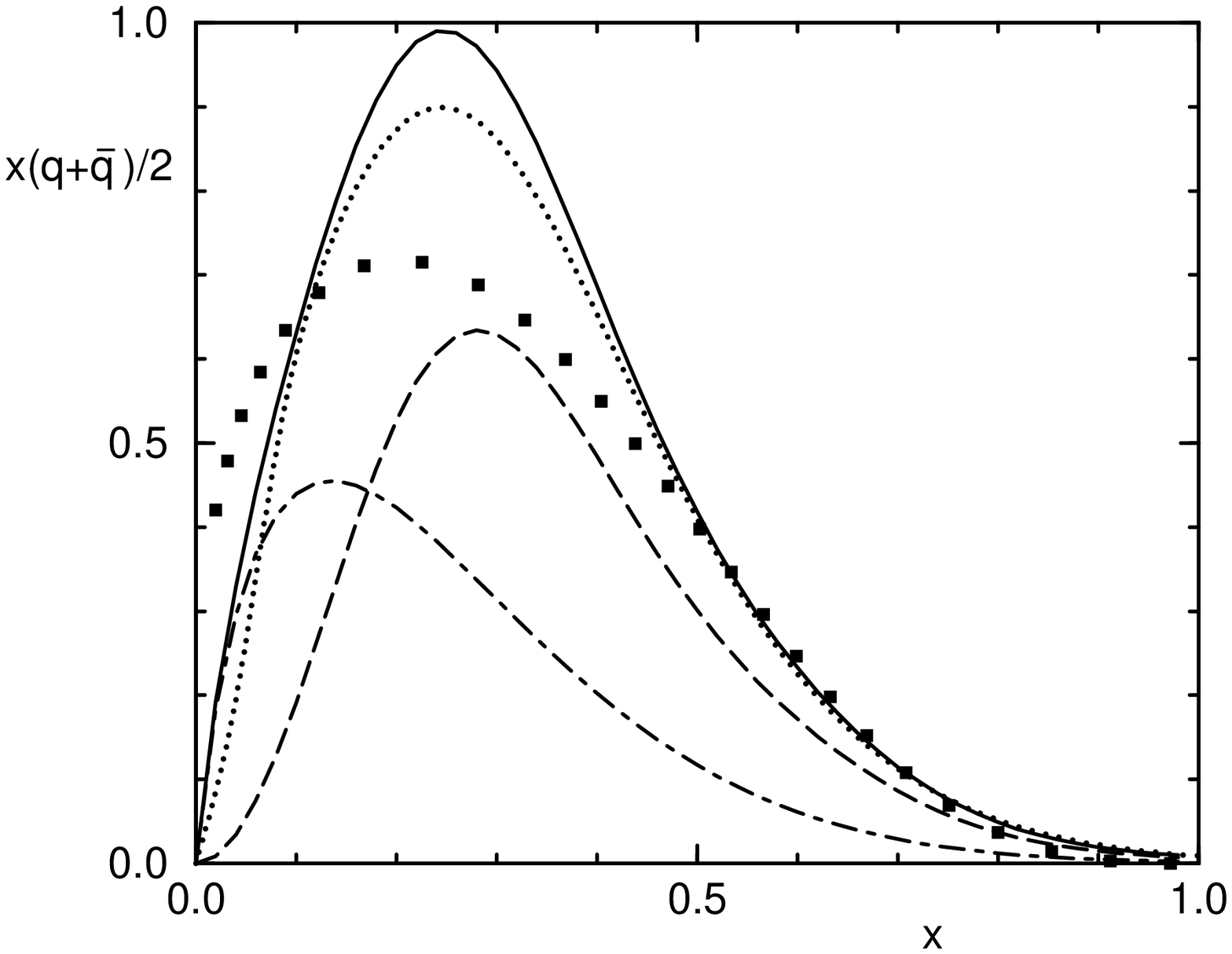}}
\caption{The singlet unpolarized distribution, $x[u(x)+d(x)+\bar
u(x)+\bar d(x)]/2$. Dashed line: regularized contribution from the discrete
level; dash-dotted line: contribution from the Dirac continuum according
to the interpolation formula, \protect\eq{interpol}; solid line: the total
distribution being the sum of the dashed and dash-dotted curves, dotted
line: the exact total distribution; squares: the parametrization of
ref.~\protect\cite{GRV}.}
\end{figure}

\newpage
\begin{figure}
 \vspace{-1cm}
\epsfxsize=16cm
\epsfysize=15cm
\centerline{\epsffile{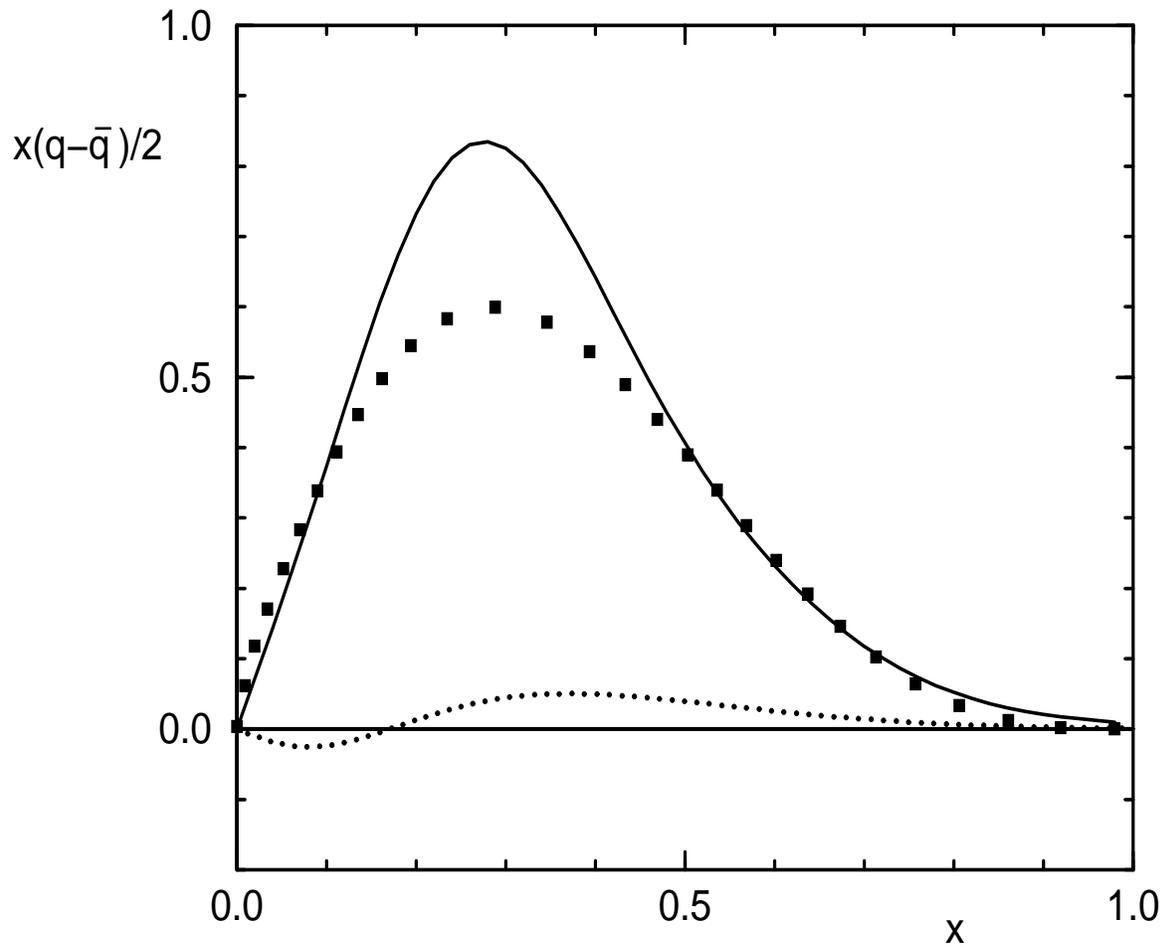}}
\caption[]{The baryon number distribution, $x[u(x)+ d(x)
-\bar u(x)-\bar d(x)]/2$. Solid line:  distribution from the unregularized
discrete level, \protect\eq{valence1}; dotted line: exact Dirac continuum
contribution; squares: the parametrization of ref.~\protect\cite{GRV}.}
\end{figure}

\newpage
\begin{figure}
 \vspace{-1cm}
\epsfxsize=16cm
\epsfysize=15cm
\centerline{\epsffile{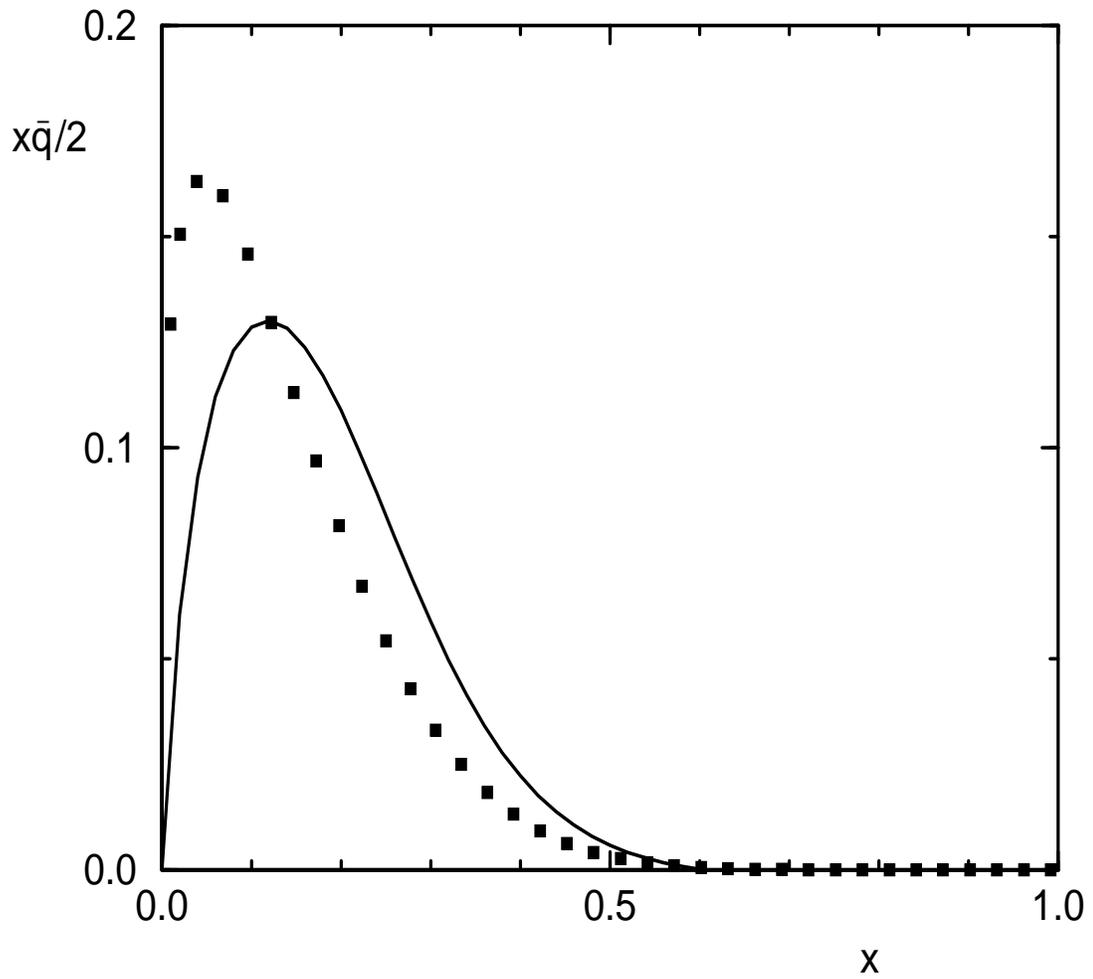}}
\caption{The antiquark distribution, $x[\bar u(x)+\bar d(x)]/2$. Solid
line: theory; squares: the parametrization of ref.~\protect\cite{GRV}.}
\end{figure}

\newpage
\begin{figure}
 \vspace{-1cm}
\epsfxsize=16cm
\epsfysize=15cm
\centerline{\epsffile{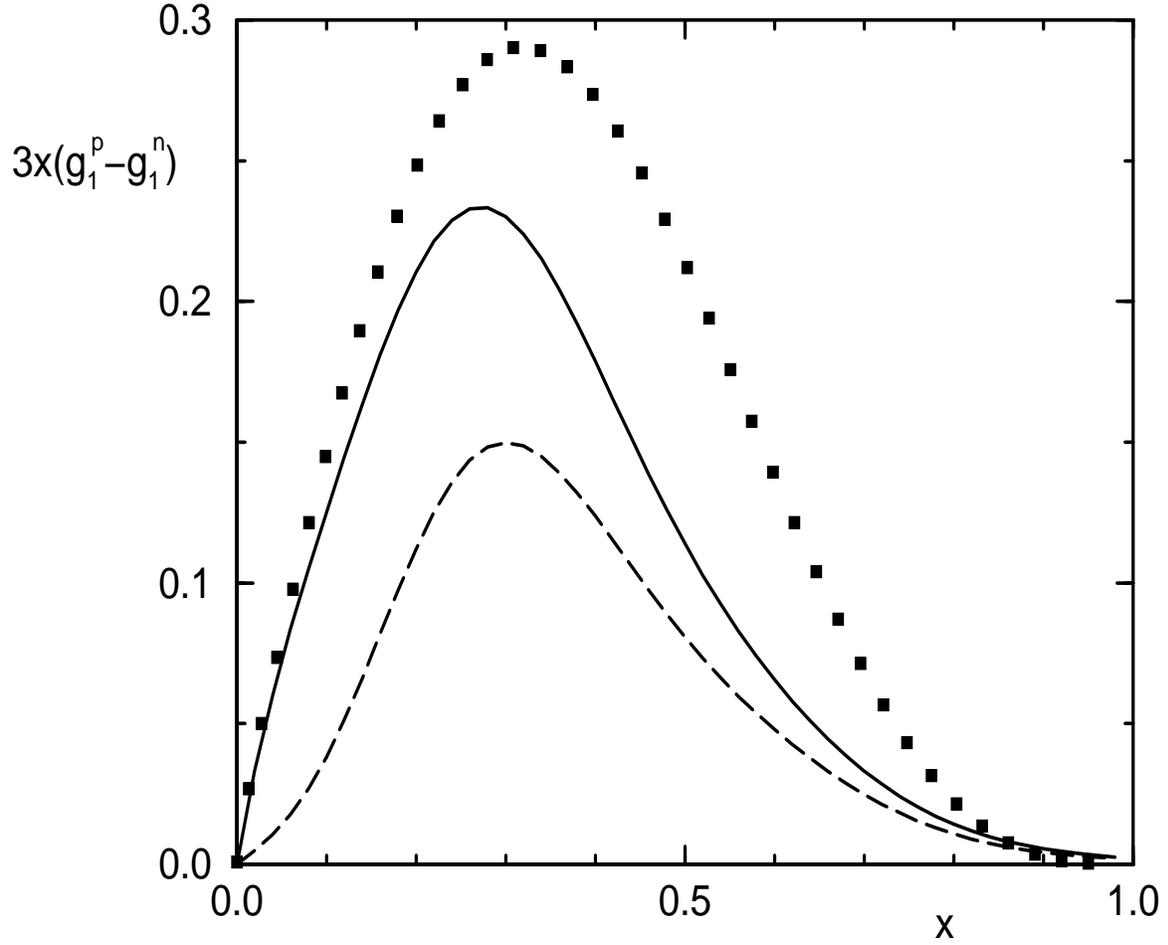}}
\caption[]{The isovector polarized distribution,
$x[\Delta u(x)-\Delta d(x)+\Delta\bar u(x)-\Delta\bar d(x)]/2$.
Dashed line: regularized contribution from the discrete level; solid
line: the sum of the contributions from the discrete level and
from the continuum according to \eq{polinterpol}; squares:  the
parametrization of ref.~\protect\cite{GR2}.}
\end{figure}

\newpage
\begin{figure}
 \vspace{-1cm}
\epsfxsize=16cm
\epsfysize=15cm
\centerline{\epsffile{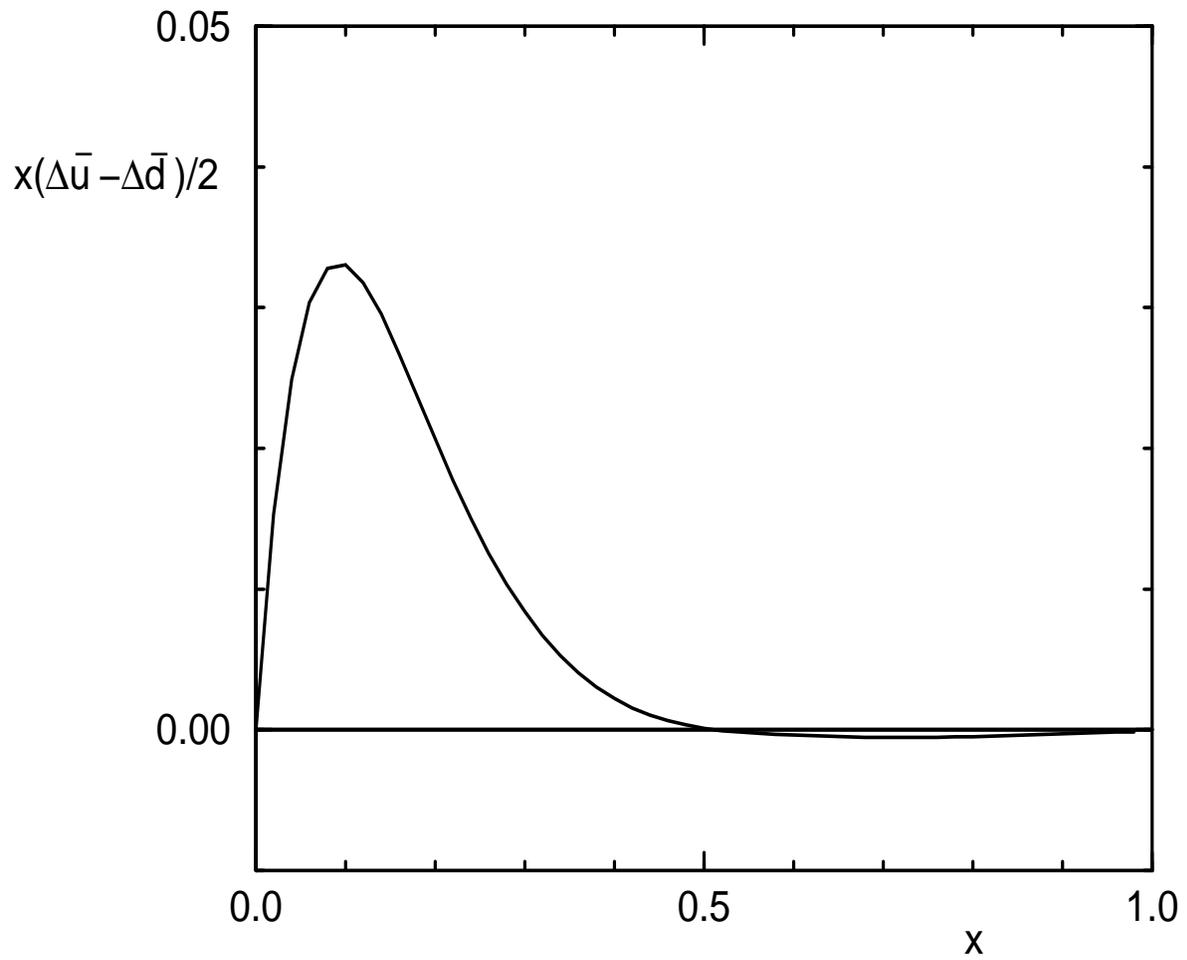}}
\caption[]{The isovector polarized distribution of antiquarks,
$x[\Delta\bar u(x)-\Delta\bar d(x)]/2$. Ref.~\protect\cite{GR2} assumes
this quantity to be zero.}
\end{figure}

\end{document}